\documentclass[aps,pra,onecolumn,superscriptaddress,groupedaddress]{revtex4}  
\usepackage{graphicx}  
\usepackage{dcolumn}   
\usepackage{bm}        
\usepackage{amsthm}
\usepackage{amsmath}
\usepackage{amssymb}
\usepackage{amsfonts}
\usepackage{amscd}
\usepackage{enumerate}
\usepackage{bm}
\usepackage{latexsym}
\usepackage{color}
\usepackage{hyperref}


\def\;{{\hspace{0.3ex};\hspace{0.5ex}}}
\def\,{{\hspace{0,3ex},\hspace{0.5ex}}}
\def\({{\hspace{1.2ex}(}}

\def\lin{\mathop{\rm span}}

\def\dim{{{\rm dim}}}
\def\rank{{{\rm rank}}}

\def\QED{\mbox{\rule[0pt]{1.5ex}{1.5ex}}}

\def\endproof{\hspace*{\fill}~\QED\par\endtrivlist\unskip}

\newtheorem{definition}{Definition}
\newtheorem{proposition}[definition]{Proposition}
\newtheorem{lemma}[definition]{Lemma}

\newtheorem{theorem}[definition]{Theorem}
\newtheorem{corollary}[definition]{Corollary}
\newtheorem{conjecture}[definition]{Conjecture}

\newtheorem{remark}[definition]{Remark}
\newtheorem{example}[definition]{Example}
\newtheorem{question}[definition]{Question}

\newcommand{\nc}{\newcommand}

 \nc{\bbA}{\mathbb{A}} \nc{\bbB}{\mathbb{B}} \nc{\bbC}{\mathbb{C}}
 \nc{\bbD}{\mathbb{D}} \nc{\bbE}{\mathbb{E}} \nc{\bbF}{\mathbb{F}}
 \nc{\bbG}{\mathbb{G}} \nc{\bbH}{\mathbb{H}} \nc{\bbI}{\mathbb{I}}
 \nc{\bbJ}{\mathbb{J}} \nc{\bbK}{\mathbb{K}} \nc{\bbL}{\mathbb{L}}
 \nc{\bbM}{\mathbb{M}} \nc{\bbN}{\mathbb{N}} \nc{\bbO}{\mathbb{O}}
 \nc{\bbP}{\mathbb{P}} \nc{\bbQ}{\mathbb{Q}} \nc{\bbR}{\mathbb{R}}
 \nc{\bbS}{\mathbb{S}} \nc{\bbT}{\mathbb{T}} \nc{\bbU}{\mathbb{U}}
 \nc{\bbV}{\mathbb{V}} \nc{\bbW}{\mathbb{W}} \nc{\bbX}{\mathbb{X}}
 \nc{\bbZ}{\mathbb{Z}}

\def\diag{\mathop{\rm diag}}

\nc{\cA}{{\cal A}} \nc{\cB}{{\cal B}} \nc{\cC}{{\cal C}}
\nc{\cD}{{\cal D}} \nc{\cE}{{\cal E}} \nc{\cF}{{\cal F}}
\nc{\cG}{{\cal G}} \nc{\cH}{{\cal H}} \nc{\cI}{{\cal I}}
\nc{\cJ}{{\cal J}} \nc{\cK}{{\cal K}} \nc{\cL}{{\cal L}}
\nc{\cM}{{\cal M}} \nc{\cN}{{\cal N}} \nc{\cO}{{\cal O}}
\nc{\cP}{{\cal P}} \nc{\cQ}{{\cal Q}} \nc{\cR}{{\cal R}}
\nc{\cS}{{\cal S}} \nc{\cT}{{\cal T}} \nc{\cU}{{\cal U}}
\nc{\cV}{{\cal V}} \nc{\cW}{{\cal W}} \nc{\cX}{{\cal X}}
\nc{\cZ}{{\cal Z}}

\def\a{\alpha}
\def\b{\beta}
\def\g{\gamma}
\def\d{\delta}

\def\p{\pi}
\def\r{\rho}

\def\ps{\psi}
\def\o{\omega}

\def\dg{\dagger}

\def\op{\oplus}
\def\ox{\otimes}

\newcommand{\bra}[1]{\langle#1|}
\newcommand{\ket}[1]{|#1\rangle}
\newcommand{\proj}[1]{| #1\rangle\!\langle #1 |}
\newcommand{\ketbra}[2]{|#1\rangle\!\langle#2|}
\newcommand{\braket}[2]{\langle#1|#2\rangle}
\newcommand{\abs}[1]{|#1|}

\nc{\U}{\mathrm{U}}

\def\bcj{\begin{conjecture}}
\def\ecj{\end{conjecture}}
\def\bcr{\begin{corollary}}
\def\ecr{\end{corollary}}
\def\bd{\begin{definition}}
\def\ed{\end{definition}}
\def\bea{\begin{eqnarray}}
\def\eea{\end{eqnarray}}
\def\bem{\begin{enumerate}}
\def\eem{\end{enumerate}}
\def\bex{\begin{example}}
\def\eex{\end{example}}
\def\bim{\begin{itemize}}
\def\eim{\end{itemize}}
\def\bl{\begin{lemma}}
\def\el{\end{lemma}}
\def\bpf{\begin{proof}}
\def\epf{\end{proof}}
\def\bpp{\begin{proposition}}
\def\epp{\end{proposition}}
\def\bqu{\begin{question}}
\def\equ{\end{question}}
\def\br{\begin{remark}}
\def\er{\end{remark}}
\def\bt{\begin{theorem}}
\def\et{\end{theorem}}

\begin{document}


\title{Product states and Schmidt rank of mutually unbiased bases in dimension six}

\author{Lin Chen}\email{linchen@buaa.edu.cn}
\affiliation{School of Mathematics and Systems Science, Beihang University, Beijing 100191, China}
\affiliation{International Research Institute for Multidisciplinary Science, Beihang University, Beijing 100191, China}
\author{Li Yu}\email{yupapers@sina.com}
\affiliation{Department of Physics, Hangzhou Normal University, Hangzhou, Zhejiang 310036, China}
\affiliation{National Institute of Informatics, 2-1-2 Hitotsubashi, Chiyoda-ku, Tokyo 101-8430, Japan}

\date{\today}

\begin{abstract}
We show that if a set of four mutually unbiased bases (MUBs) in $\mathbb{C}^6$ exists and contains the identity, then any other basis in the set contains at most two product states and at the same time has Schmidt rank at least three. Here both the product states and the Schmidt rank are defined over the bipartite space $\mathbb{C}^2\otimes\mathbb{C}^3$. We also investigate the connection of the Sinkhorn normal form of unitary matrices to the fact that there is at least one vector unbiased to any two orthonormal bases in any dimension.
\end{abstract}




\maketitle

\tableofcontents

\section{Introduction}

The concept of complementary observables is ubiquitous in quantum physics. The observables are described using Hermitian operators. Some pairs of unitary operators can also be regarded as being complementary to each other, as mentioned in Schwinger's work \cite{schwinger60} in 1960. These two concepts have a common definition as follows. The eigenvectors of two complementary (Hermitian or unitary) operators form two bases in the complex Hilbert space $\bbC^d$ satisfying the following condition: the modulus of the inner product of any two vectors respectively from the two bases is ${1\over\sqrt d}$. Such two vectors are mutually unbiased (MU), and the two bases are MU bases (MUBs). The MUBs have various applications in quantum tomography, cryptography, and the construction of Wigner functions. Some of these applications are explained in Brierley's PhD thesis \cite{mub09}. For example, a complete set of $d+1$ MUBs in $d$ dimensions is optimal for minimizing the statistical uncertainty in estimating a density matrix. And sets of MUBs could be used to hide information from an eavesdropper in a quantum key distribution protocol. The connection to discrete Wigner functions is illustrated in \cite{Assw14} (with errata \cite{Assw15}), which discusses the proporties of states that yield the same probability distributions under measurements corresponding to a set of MUBs.

In this paper, we identify a unitary matrix with an orthonormal basis consisting of the column vectors of the matrix. We say that $n$ order-$d$ unitary matrices form $n$ MUBs if the inner product of any two column vectors from different matrices is a complex number of modulus $1/\sqrt d$ \cite{schwinger60}. An order-$d$ unitary matrix whose elements all have modulus $1/\sqrt d$ is a \textit{complex Hadamard matrix (CHM)}. It is known that a triple of MUBs in $\bbC^6$ have been constructed. It is conjectured that
\bcj
\label{cj:mub6}
Four MUBs in $\bbC^6$ do not exist.
\ecj
We refer the readers to recent progress on MUBs in \cite{bw08,bw09,bw10,mb15}. Conjecture \ref{cj:mub6} is an open problem in quantum physics and quantum information. In this paper we investigate this conjecture in terms of the product vectors and Schmidt rank of matrices introduced in Sec. \ref{sec:pre}. It has been proved that any set of seven MUBs in $\bbC^6$ contains at most one product-vector basis \cite{mw12jpa102001}, where the product-vector basis is defined in Sec.~\ref{subsec:equivalence} below. After looking into the paper \cite{jmm09} cited by \cite{mw12jpa102001}, the ``seven'' above can be replaced with ``four''. We shall exclude the existence of a few families of CHMs in sets of four MUBs in $\bbC^6$ containing the identity matrix. (Throughout the paper, ``to exclude something'' means to eliminate the possibility that something may appear or exist. For example, when we say a case is ``excluded'' in a proof, we mean that this case is impossible, and not just that we do not discuss this case.) Our first main result is as follows.


\bt
\label{thm:2prod}
If a set of four MUBs in $\bbC^6$ contains the identity, then
\\
(i) any other MUB in the set contains at most two product column vectors.
\\
(ii) the other three MUBs in the set contains totally at most six product column vectors.
\et
We will prove assertion (i) by investigating the matrix $Y_4$ in Lemma \ref{le:mub} (v), and assertion (ii) follows from (i). Theorem \ref{thm:2prod} restricts the number of product column vectors in a set of four MUBs in $\bbC^6$. For example, the well-known Fourier matrix $F_6$ of order six contains six product column vectors, and thus it cannot be a member of any four MUBs containing the identity in $\bbC^6$. Here
\bea
\label{eq:f6}
F_6=
{1\over\sqrt6}\left(
                   \begin{array}{cccccc}
                     1 & 1 & 1 & 1 & 1 & 1\\
                     1 & \a & \a^2 & \a^3 & \a^4 & \a^5\\
                     1 & \a^2 & \a^4 & \a^6 & \a^8 & \a^{10}\\
                     1 & \a^3 & \a^6 & \a^9 & \a^{12} & \a^{15}\\
                     1 & \a^4 & \a^8 & \a^{12} & \a^{16} & \a^{20}\\
                     1 & \a^5 & \a^{10} & \a^{15} & \a^{20} & \a^{25}\\
                     \end{array}
                 \right)
\eea
where $\a=e^{\p i \over 3}$. The subset of a complete set of MUBs is defined as an MU constellation \cite{mub09}. It is known that an MU constellation consisting of three MUBs plus one additional vector, related to the Heisenberg-Weyl group, does not exist \cite{mub09}. So the constellation of such type has at most 18 product vectors. In contrast, Theorem \ref{thm:2prod} (ii) shows that any family of MUB constellation of four MUBs containing the identity matrix has at most 12 product vectors.

We further show that a set of four MUBs containing the identity matrix cannot also contain a matrix with an order-three submatrix proportional to a unitary matrix or a real submatrix of size $3\times2$. This is proved by the matrices $Y_1$ and $Y_6$ in Lemma \ref{le:mub} (v). It excludes the possibility that some known CHMs may be the members of four MUBs containing the identity matrix. They include the Dita matrix \cite[Eq. (C.1)]{mub09}, the spectral matrix \cite[Eq. (C.4)]{mub09}, and some non-affine CHMs such as the symmetric family \cite[Eq. (C.9)]{mub09}.

As an application to entanglement theory, we will apply Theorem \ref{thm:2prod} to construct a family of $2\times3$ entangled states in Lemma \ref{le:entanglement}.

Next, we introduce our second main result.
\bt
\label{thm:sr3}
If a set of four MUBs in $\bbC^6$ contains the identity, then any other MUB in the set has Schmidt rank at least three.
\et
The Schmidt rank is defined in Sec.~\ref{sec:sr}, and the proof is at the end of Sec. \ref{subsec:mubtrio}. Theorem \ref{thm:sr3} excludes the possibility that some known CHMs may be the members of four MUBs containing the identity matrix. They include the Bjorck's circulant matrix in \cite[Eq. (C.2)]{mub09}, see Sec.~\ref{subsec:mubtrio} for details.
To understand how the properties of composite dimensions and prime dimensions are different from each other is a main motivation in \cite{bw08}, and our Theorem \ref{thm:sr3} can be viewed as a step towards the full development of such motivation. In the following we introduce the physical meanings of the Schmidt rank and how they are related to the CHMs.

It is known that any $2\times3$ bipartite unitary operator of Schmidt rank three is a controlled unitary operator \cite{cy14}.
The Schmidt rank plays an important role in quantum computing and has received extensive research in recent years. Especially, they have been used to evaluate the following three quantities: entangling power, assisted entangling power and disentangling power of bipartite unitaries  \cite{cy13,cy14,cy14ap,cy16}. The first two quantities quantitatively characterize the maximum amount of entanglement increase when the input states are respectively a product state and an arbitrary pure state. The third quantity describes the maximum amount of entanglement decrease over all pure input states. The maximum amount of entanglement increase over all input states is a lower bound of the entanglement cost for implementing bipartite unitaries under local operations and classical communication (LOCC). Since a CHM in composite dimension can be viewed as a special bipartite unitary operator, we hope the introduction of the Schmidt rank could further the study of CHMs, and that in turn would help the study of MUBs.

The rest of this paper is organized as follows. In Sec. \ref{sec:pre} we introduce in Definition \ref{df:mub} the notations used in the paper, such as the Schmidt rank of bipartite unitary operators, equivalent MUBs, the MUB trio and some results from linear algebra. We further construct the preliminary results in Lemma \ref{le:van} and \ref{le:la}. In Sec. \ref{sec:result} we construct results for order-six MUBs and complex Hadamard matrices in Lemma \ref{le:chm}. We show that there is at least one vector unbiased to any two given orthonormal bases in any dimension in Lemma \ref{le:mubsinkhorn}, which is a direct consequence of the Sinkhorn normal form of unitary matrices proved in \cite{iw15}. We shall exclude the existence of several types of order-six CHM as a member of some MUB trio in Lemma \ref{le:mub}. We further apply Theorem \ref{thm:2prod} to construct Lemma \ref{le:entanglement}.
 Using the above-mentioned lemmas we construct Lemma \ref{le:sr2} as a preliminary result for Theorem \ref{thm:sr3} and Corollary \ref{cr:sr3}. We also carry out a few numerical studies about some CHMs in Sec.~\ref{ssec:num}. We propose a few open problems in Sec. \ref{sec:pro}.

\section{Preliminaries}
\label{sec:pre}
	
In this section we introduce the notations and preliminary results used in the paper. They include the Schmidt rank of bipartite unitary operators and controlled unitary operators in Sec. \ref{sec:sr}, unitary equivalence and local equivalence of MUBs in Definition \ref{df:mub} in Sec. \ref{subsec:equivalence}, and linear algebra in Lemmas \ref{le:van} and \ref{le:la} in Sec. \ref{subsec:la}. In particular, Lemma \ref{le:la} consists of results on the product states, matrix rank, permutation matrices, and so on. While useful for the current topic of MUBs, these results have broader applications in quantum information, such as the separable and positive-partial-transpose (PPT) states.

\subsection{Schmidt rank and controlled unitary operators}
\label{sec:sr}

We denote $\ket{i,j},i=1,\cdots,d_A$, $j=1,\cdots,d_B$ as the computational-basis states of the bipartite Hilbert space $\cH=\cH_A\ox\cH_B=\bbC^{d_A}\ox\bbC^{d_B}$. Let
$I_d$ be the identity matrix of order $d$. The bipartite unitary gate $U$ acting on $\cH$ has \emph{Schmidt rank} $n$ if there is an expansion $U=\sum^n_{j=1}A_j \ox B_j$ where the order-$d_A$ matrices $A_1,\cdots,A_n$ are linearly independent, and the order-$d_B$ matrices $B_1,\cdots,B_n$ are also linearly independent. The Schmidt rank is equivalent to the notion of operator-Schmidt rank in \cite{Nielsen03,Tyson03}.

We say that $U$ is a \textit{controlled unitary gate}, if $U$ is equivalent to $\sum^{d_A}_{j=1}\proj{j}\ox U_j$ or
$\sum^{d_B}_{j=1}V_j \ox \proj{j}$ via local unitaries. Further, $U$ is controlled in the computational basis from $A$ side if $U=\sum^{d_A}_{j=1}\proj{j}\ox U_j$. Bipartite unitary gates of Schmidt rank two or three are in fact controlled unitaries \cite{cy13,cy14,cy14ap}.
A permutation matrix (or equivalently a permutation unitary) is a unitary matrix containing elements $0$ and $1$ only.
A complex permutation matrix is a unitary matrix with one and only one nonzero element in each row and column.

\subsection{The equivalence of MUBs}
\label{subsec:equivalence}

To investigate Conjecture \ref{cj:mub6}
we review the following definitions. When $d=pq$ where $p,q>1$, we call a basis of $\bbC^d$ consisting of product vectors in $\bbC^p\ox\bbC^q$ as a product-vector basis. We call a square matrix as a product-vector matrix when its columns form a product-vector basis. We say that $n$ unitary matrices form $n$ product-vector MUBs when these matrices are all product-vector matrices and they form $n$ MUBs.  For a square matrix $C$, we say $C$ is a direct-product matrix if $C=F\ox G$ where $F$ and $G$ are square matrices of order greater than one (in this paper the orders of $F$ and $G$ are fixed when we use such definition). Our definition of product-vector basis corresponds to both the direct product basis and the indirect product basis in \cite{wpz11}.  We refer to the \textit{subunitary matrix} as the matrix proportional to a unitary matrix. The $H_2$-reducible CHM is an order-six CHM that contains a subunitary submatrix of order two \cite{karlsson11}. Now we present the following definitions.
\bd
\label{df:mub}
(i) Let $U_1,\cdots,U_n$ be $n$ unitary matrices of order $d$. They form $n$ MUBs if and only if for an arbitrary unitary matrix $X$, and arbitrary complex permutation matrices $P_1,\cdots,P_n$, the $n$ matrices $XU_1P_1,\cdots,XU_nP_n$ form $n$ MUBs.
In this case we say that $U_1,\cdots,U_n$ and $XU_1P_1,\cdots,XU_nP_n$ are \textbf{unitarily equivalent} MUBs. Furthermore they are \textbf{locally unitarily (LU) equivalent} MUBs when $X$ is a direct-product matrix.

Let $U_1,\cdots,U_n$ be product-vector MUBs such that $U_j=(\cdots,\ket{a_{jk},b_{jk}},\cdots)$ where $\ket{a_{jk}}\in\bbC^p$ and $\ket{b_{jk}}\in\bbC^q$. Let
$U_j^{\Gamma_A}$ and $U_j^{\Gamma_B}$ both denote $U_j$ except that $\ket{a_j}$ and $\ket{b_j}$ are respectively replaced by their complex conjugates. Then we say that any two of the following four sets
\bea
&&
U_1,\cdots,U_n,
\\&&
U_1^{\Gamma_A},\cdots,U_n^{\Gamma_A},
\\&&
U_1^{\Gamma_B},\cdots,U_n^{\Gamma_B},
\\&&
XU_1P_1,\cdots,XU_nP_n,
\eea
are LU-equivalent product-vector MUBs, where $X$ is a direct-product matrix.
\\
(ii) Let $U,V$ and $W$ be three CHMs of order six. Conjecture \ref{cj:mub6} is equivalent to ask whether $I,U,V$ and $W$ can form four MUBs, i.e., whether $U^\dg V$, $V^\dg W$ and $W^\dg U$ are still CHMs. If they do, then we denote the set of $U,V$ and $W$ as an MUB trio.
\\
(iii) We say that two CHMs $X$ and $Y$ are \textbf{equivalent} when there exist two complex permutation matrices $C$ and $D$ such that $X=CYD$. For simplicity we refer to $X$ as $Y$ up to equivalence. The equivalence class of $X$ is the set of all CHMs which are equivalent to $X$. The minimum Schmidt rank in the equivalence class of $X$ is called the \textbf{min-Schmidt rank} of $X$.
\\(iv) In (iii), we say that $X$ and $Y$ are \textbf{locally equivalent} when $C$ is a direct-product matrix.
\ed

We say that an order-$n$ CHM is in the \textit{dephased form} when all elements in the first row and first column of the CHM are equal to $1/\sqrt n$. Evidently every CHM is equivalent to another CHM in the dephased form.

The CHMs in $\bbC^6$ have been extensively introduced in the appendix A of \cite{bw09}. In particular, it has been shown that the pair $\{I,S\}$ cannot be extended to a triple of MUBs \cite{bw09}. Here $I$ is the identity matrix, and $S$ is a CHM known as the \emph{spectral matrix} \cite{tao04}
\bea
\label{eq:spec}
S=
{1\over\sqrt6}
\left(
                   \begin{array}{cccccc}
                     1 & 1 & 1 & 1& 1& 1 \\
                     1 & 1 & \o & \o& \o^2& \o^2 \\
                     1 & \o & 1 & \o^2& \o^2& \o \\
                     1 & \o & \o^2 & 1& \o& \o^2 \\
                     1 & \o^2 & \o^2 & \o& 1& \o \\
                     1 & \o^2 & \o & \o^2& \o& 1 \\
                     \end{array}
                 \right),
\eea
where $\o:=e^{2\p i/3}$.
We will also encounter the following class of unitaries $U$ which is the standard form for CHMs in the \emph{Fourier family} $F_6^{(2)}$ (c.f. \cite[Sec. 6]{karlsson11}):
\bea\label{eq:ufourier1}
U=
\frac{1}{\sqrt{6}}
\left(
                   \begin{array}{cccccc}
                     1 & 1 & 1 & 1 & 1 & 1 \\
                     1 & -1 & z_1 & -z_1 & z_2 & -z_2 \\
                     1 & 1 & \o & \o & \o^2 & \o^2 \\
                     1 & -1 & \o z_1 & -\o z_1 & \o^2 z_2 & -\o^2 z_2 \\
                     1 & 1 & \o^2 & \o^2 & \o & \o \\
                     1 & -1 & \o^2 z_1 & -\o^2 z_1 & \o z_2 & -\o z_2 \\
                     \end{array}
                 \right),
\eea
where $\vert z_1\vert=\vert z_2\vert=1$.
Any other member in the Fourier family $F_6^{(2)}$ is equivalent to the form above.
The \emph{Fourier transposed family} $(F_6^{(2)})^T$ (c.f. \cite[Sec. 6]{karlsson11}) consists of the transpose of the CHMs in $F_6^{(2)}$.

From now on we regard any order-six CHM as a $2\times3$ bipartite unitary. One may easily check that $S$ has Schmidt rank four. The general CHM $U$ has Schmidt rank at most four and we exemplify the $U$'s with arbitrary Schmidt rank. First,
$
U={1\over\sqrt2}
\left(
                   \begin{array}{cccccc}
                     1 & 1 \\
                     1 & -1 \\
                     \end{array}
                 \right)
\ox
{1\over\sqrt3}
\left(
                   \begin{array}{cccccc}
                     1 & 1 & 1\\
                     1 & \o & \o^2 \\
                     1 & \o^2 & \o \\
                     \end{array}
                 \right)
$ has Schmidt rank one.
Next,
$U=
{1\over\sqrt2}
\left(
                   \begin{array}{cccccc}
                     V & W \\
                     V & -W \\
                     \end{array}
                 \right)$ has Schmidt rank two
when $V$ and $W$ are two CHMs of order three. Another Schmidt-rank-two CHM is the so-called Björck’s circulant matrix, see \cite[Eq. (A2)]{bw09}. It has the expression $
\left(
                   \begin{array}{cccccc}
                     X & Y \\
                     Y & X \\
                     \end{array}
                 \right),$ and neither of the order-three matrices $X$ and $Y$ is proportional to a unitary matrix.
Third,
\bea
\label{eq:hadamard=sr3}
U={1\over\sqrt2}
\left(
                   \begin{array}{cccccc}
                     \diag(-i,i,1)\cdot V &  V \\
                       V &  \diag(-i,i,-1)\cdot V \\
                     \end{array}
                 \right)
\eea
is a CHM of Schmidt rank three, where $V$ is a CHM of order three.
$U$ may also be of Schmidt rank four, e.g., the spectral matrix in \eqref{eq:spec}. For any Schmidt-rank four $U$, we may assume that
\bea
U
=\sum^2_{j,k=1} \ketbra{j}{k} \ox U_{jk}
={1\over2}\sum^2_{j,k=1} (\sqrt2\ketbra{j}{k}) \ox (\sqrt2 U_{jk}).
\eea
One may check that each of $\{(\sqrt2\ketbra{j}{k}\}$ and $\{\sqrt2 U_{jk}\}$ is a linearly independent set, and the former is orthonormal.
One may easily extend the above examples to $2\times d_B$  CHMs $U$ of Schmidt rank two, three and four. We construct an example of Schmidt-rank-four. Let
\bea
U={1\over\sqrt2}
\left(
                   \begin{array}{cccccc}
                     D\cdot V & D \cdot W \\
                     V & -W \\
                     \end{array}
                 \right)
\eea
where $D,V,W$ are all order-$d_B$ CHMs, $D$ is a diagonal unitary and not proportional to the identity matrix, $V,W$ are linearly independent and $WV^\dg$ is not diagonal. Then one can show that $U$ is a Schmidt-rank-four $2\times d_B$ CHM.

\subsection{Linear algebra}
\label{subsec:la}

For studying Conjecture \ref{cj:mub6}, we propose a few preliminary facts on linear algebra in Lemma \ref{le:la}. First of all we introduce a result from \cite[Lemma 1]{cy16} about diagonal unitary matrices. In this paper we will apply it to vectors, since the set of diagonal unitary matrices of order $d$ is isomorphic to the set of $d$-dimensional column vectors with elements of modulus $1/\sqrt d$.

\bl
\label{le:van}
Let $D$ be a diagonal unitary matrix. The following four statements are equivalent.
\\
(i) $D$ has at least three distinct eigenvalues;
\\
(ii) the identity, $D$ and $D^\dg$ are linearly independent;
\\
(iii) any unitary in the linear span of the identity and $D$ is proportional to one of them;
\\
(iv) any multiple of unitary in the linear span of the identity and $D$ is proportional to one of them.
\qed
\el
For further investigation, we introduce a few definitions.
The \textit{generalized permutation matrix (GPM)} is a square matrix such that there is exactly one nonzero entry in each row and each column of  the matrix. For example the complex permutation matrix is a GPM. We shall refer to the \textit{doubly quasistochastic matrices} as the matrices whose every row's and column's elements sum to one \cite[Definition 5.11]{idel16}. Such matrices reduce to the doubly stochastic matrices when they are square matrices and of all nonnegative elements. We further denote $\r^\Gamma$ as the partial transpose w.r.t. the first system of the bipartite state $\r$, and $\r$ is PPT when $\r^\Gamma\ge0$.

\bl
\label{le:la}
(i) Suppose $1+e^{i\a}+e^{i\b}=0$, where $\a,\b\in[0,2\p)$. Then $(\a,\b)=( {2\over3}\p,{4\over3}\p)$ or $( {4\over3}\p,{2\over3}\p)$.
\\
(ii) Suppose $[u_{jk}]$ is an order-three unitary matrix of at most one zero entry, and  $[p_{jk}u_{jk}]$ is also a unitary matrix where $p_{11}=p_{12}=p_{13}=p_{21}=p_{31}=1$, and $p_{jk}=\pm 1$ otherwise. Then $[p_{jk}u_{jk}]=D_1 [u_{jk}] D_2$ where $D_1$ and $D_2$ are diagonal real unitary matrices.
\\
(iii) If an orthonormal basis in $\bbC^2\ox\bbC^3$ contains five product states, then the remaining state of the basis is also a product state.
\\
(iv) Suppose $e^{ia}+e^{ib}=e^{ic}+e^{id}$ where $a,b,c,d\in[0,2\p)$. Then we have either of the three cases: $a=c,b=d$, or $a=d,b=c$, or $\abs{a-b}=\abs{c-d}=\p$.
\\
(v) Suppose $\left(
                   \begin{array}{cccccc}
                     V_0 & V_1 \\
                     V_2 & V_3 \\
                     \end{array}
                 \right)$ is a partitioned unitary matrix of order $d$. Then $V_0$ and $V_3$ have full rank at the same time. Furthermore if $V_0$ and $V_3$ are both square matrices, then $\rank V_0 + \rank V_3 \le d-2\dim\ker V_0$.
\\                 
(vi) Let $u, v, s, t$ be complex numbers of modulus one. Then $(u+v) (s^* + t^* ) ( u^* s + v^* t )$ is real.
\\
(vii) For any unitary matrix $U$, there are diagonal unitaries $L$ and $R$ such that $L U R$ is doubly quasistochastic. Neither $L$ or $R$ are necessarily unique.
\\
(viii) For any positive definite matrix $W$, there are diagonal matrices $L$ and $R$ such that $L W R$ is doubly quasistochastic. Neither $L$ nor $R$ are necessarily unique.
\\
(ix) Suppose an orthonormal basis in $\bbC^6$ contains $k$ product states. Then the remaining $6-k$ states in the basis span a subspace spanned by orthogonal product vectors.
\\
(x) Let $U$ be a bipartite unitary matrix. Then the min-Schmidt rank of $U$ is equal to those of $U^\dag$, $U^T$ and $U^\ast$.
\el
We prove Lemma \ref{le:la} in Appendix \ref{app:{le:la}}. Note that one of the diagonal matrices $L$ and $R$ in (viii) may be not unitary. An example is $W=\diag(2,1)$. Whether one of $L$ and $R$ can be chosen as a unitary matrix for any $W$ is unknown.

\section{Results}
\label{sec:result}

In this section we introduce our main results. In Sec. \ref{subsec:chm} we investigate the order-six CHMs in terms of the subunitary matrices in Lemma \ref{le:chm}. In Sec. \ref{subsec:mub}, we first show that there is at least one vector unbiased to any two given orthonormal bases in any dimension in Lemma \ref{le:mubsinkhorn}, which is a direct consequence of the Sinkhorn normal form of unitary matrices proved in \cite{iw15}. We discuss the number of such vectors in Conjecture~\ref{cj:muvectors} and Lemma~\ref{lm:muvectors}. We further construct a few preliminary results on MUBs in Lemma \ref{le:mub}. They show that some CHMs do not exist in four MUBs containing the identity matrix. In Sec. \ref{subsec:mubtrio} we show that the member of any MUB trio has Schmidt rank at least three in Theorem \ref{thm:sr3}. This is based on the preliminary Lemma \ref{le:sr2}. We also show in Corollary \ref{cr:sr3} that if a set of four MUBs in $\bbC^6$ contains the identity, then any other MUB in the set has min-Schmidt rank at least three. In Sec. \ref{ssec:num} we carry out a few numerical studies about the unbiased vectors, the Schmidt rank, and the submatrices of some CHMs.

\subsection{Complex Hadamard matrices}
\label{subsec:chm}

In the following lemma we provide a few properties of the order-six CHMs. For example,
we show in Lemma \ref{le:chm} (iii) that such CHMs do not contain any rank-one order-three submatrix, any order-four submatrix of rank at most two, or any singular order-five submatrix. We further reveal some relations between elements of $H_2$-reducible CHMs in (vi).

\bl
\label{le:chm}
(i) Every order-three CHM can be written as $D_1 V D_2$ where $D_1$ and $D_2$ are both diagonal unitaries, and $V={1\over\sqrt3}\left(
                   \begin{array}{cccccc}
                     1 &  1 & 1 \\
                     1 &  \o & \o^2 \\
                     1 &  \o^2 & \o \\
                     \end{array}
                 \right)$ or ${1\over\sqrt3}\left(
                   \begin{array}{cccccc}
                     1 &  1 & 1 \\
                     1 &  \o^2  & \o \\
                     1 &  \o & \o^2  \\
                     \end{array}
                 \right)$.
\\
(ii) If an order-six CHM has a submatrix of size $2\times k$ and rank one, then $k\le3$ and the equality is achievable.
\\
(iii) The order-six CHM does not have any rank-one order-three submatrix, any order-four submatrix of rank at most two, or any singular order-five submatrix.
\\
(iv) The order-six CHM is an $H_2$-reducible CHM if and only if the CHM is equivalent to another CHM containing two product column vectors $\ket{a,b},\ket{a^\perp,c}$ where $\ket{a},\ket{a^\perp}$ is an orthonormal basis in $\bbC^2$.
\\
(v) The order-six CHM has a submatrix of size $3\times2$ with orthogonal column vectors if and only if the CHM is equivalent to another CHM containing two product column vectors $\ket{a,b},\ket{c,b^\perp}$ where $\ket{b},\ket{b^\perp}$ is an orthonormal basis in $\bbC^3$.
\\
(vi) The $H_2$-reducible CHM is equivalent to the CHM $H$ in \cite[Theorem 11]{karlsson11}. We express it as
\bea
\label{eq:a12}
{1\over\sqrt6}
\left(
                   \begin{array}{cccccc}
                     1 & 1 & 1 & 1 & 1 & 1 \\
                     1 & -1 & z_1 & -z_1 & z_2 & -z_2\\
                     1 & z_3 & a_1 & a_2 & b_1 & b_2 \\
                     1 & -z_3 & a_1a_3 & -a_2a_3 & b_1b_3 & -b_2b_3 \\
                      1 & z_4 & c_1 & c_2 & d_1 & d_2 \\
                     1 & -z_4 & c_1c_3 & -c_2c_3 & d_1d_3 & -d_2d_3 \\
                     \end{array}
                 \right),
\eea
where $z_j$, $a_j$, $b_j$, $c_j$, and $d_j$ are complex numbers of modulus one and satisfy the conditions in \cite[Theorem 11]{karlsson11} and
\bea
\label{eq:z1z3}
z_1z_3&=&a_1a_2a_3,\\
\label{eq:z2z3}
z_2z_3&=&b_1b_2b_3,\\
\label{eq:z1z4}
z_1z_4&=&c_1c_2c_3,\\
\label{eq:z2z4}
z_2z_4&=&d_1d_2d_3,\\
\label{eq:a1+a3}
a_1+a_2+z_1z_3(a_2^*-a_1^*)&=&d_1+d_2+z_2z_4(d_2^*-d_1^*),\\
\label{eq:z3a1+a3}
z_3^*[a_1+a_2-z_1z_3(a_2^*-a_1^*)]&=&
z_4^*[d_1+d_2-z_2z_4(d_2^*-d_1^*)],\\
\label{eq:b1+b3}
b_1+b_2+z_2z_3(b_2^*-b_1^*)&=&c_1+c_2+z_1z_4(c_2^*-c_1^*),\\
\label{eq:z3b1+b3}
z_3^*[b_1+b_2-z_2z_3(b_2^*-b_1^*)]&=&z_4^*[c_1+c_2-z_1z_4(c_2^*-c_1^*)].
\eea
Further
\\
(vi.a) If one of the elements in the lower right order-four submatrix of \eqref{eq:a12} is equal to a constant, then up to equivalence we may assume that $a_1$ is the constant.
\\
(vi.b) Consider two matrices in \eqref{eq:a12} with $z_3=-b_1$ and $z_3=-b_2$, respectively. Then they are equivalent up to the exchange of the last two columns of \eqref{eq:a12}.
\\
(vii) Any order-six CHM does not contain an order-four or order-five subunitary submatrix.
\\
(viii) The spectral matrix in \eqref{eq:spec} is of min-Schmidt rank three. It is equivalent under complex permutation matrices to the CHM
\bea
\label{eq:spec2}
S'=
{1\over\sqrt6}
\left(
                   \begin{array}{cccccc}
                     1 & 1 & 1 & \o & 1 & \o \\
                     1 & \o & \o^2 & \o & \o^2 & \o^2 \\
                     1 & \o^2 & \o & \o^2 & \o^2 & \o \\
                     1 & \o & \o & 1 & 1 & 1 \\
                     1 & \o^2 & 1 & 1 & \o & \o^2 \\
                     \o & \o & 1 & 1 & \o^2 & \o \\
                     \end{array}
                 \right).
\eea
The upper left order-three submatrix of $S'$ is a subunitary matrix, while the upper left and bottom right order-three submatrices are equal, and the $S'$ is equivalent under local unitaries to a controlled unitary matrix.
\\
(ix) Any CHM in the Fourier family $F_6^{(2)}$ and the Fourier transposed family $(F_6^{(2)})^T$ is of min-Schmidt rank at most two.
\el

We prove Lemma \ref{le:chm} in Appendix \ref{app:{le:chm}}. In the proof of Lemma \ref{le:chm} (ii), the equality $k=3$ is also achievable when the order-six CHM $U$ is not a product matrix. An example is $\diag(1,1,1,1,1,-1)\cdot
\bigg[{1\over\sqrt2}\left(
                   \begin{array}{cccccc}
                     1 &  1 \\
                     1 &  -1  \\
                     \end{array}
                 \right)\ox {1\over\sqrt3}\left(
                   \begin{array}{cccccc}
                     1 &  1 & 1 \\
                     1 &  \o & \o^2 \\
                     1 &  \o^2 & \o \\
                     \end{array}
                 \right)\bigg]$. In spite of Lemma \ref{le:chm} (iii), the order-six CHM may contain an order-three submatrix of rank two and an order-three subunitary submatrix. An example containing both types of submatrices is ${1\over\sqrt2}\left(
                   \begin{array}{cccccc}
                     1 &  1 \\
                     1 &  -1  \\
                     \end{array}
                 \right)\ox {1\over\sqrt3}\left(
                   \begin{array}{cccccc}
                     1 &  1 & 1 \\
                     1 &  \o & \o^2 \\
                     1 &  \o^2 & \o \\
                     \end{array}
                 \right)$. We will show that such a CHM cannot be a member of any MUB trio in Lemma \ref{le:mub} (v).
Besides, Lemma \ref{le:chm} (iv) and (v) are the only two subcases of the case that a CHM contains two product column vectors. The case of Lemma \ref{le:chm} (iv), namely an order-six CHM containing an order-two subunitary matrix, has been investigated in \cite{karlsson11} and (vi). The matrices $A_{ij}$ in (vi) are subject to the constraints given in  \cite[Theorem 11]{karlsson11}. A subcase of this CHM is the Szollosi family in \cite[Eq. (C.12)]{mub09}.

Lemma~\ref{le:chm} (vii) shows that an order-six CHM does not contain an order-four or order-five subunitary submatrix. But it may contain an order-two subunitary submatrix. For example $F_6$ in \eqref{eq:f6} contains an order-two submatrix $\left(
                   \begin{array}{cccccc}
                     1 & 1 \\
                     1 & -1 \\
                     \end{array}
                 \right)$.

If $a_1=1$ in \eqref{eq:a12} then \eqref{eq:a12} contains two singular order-two submatrices. However it seems still hard to exclude \eqref{eq:a12} as a member of some MUB trio.

Next, \eqref{eq:a1+a3}-\eqref{eq:z3b1+b3} imply that the second and third order-two subunitary matrices of the bottom two rows of \eqref{eq:a12} are decided by the counterpart of the middle two rows of  \eqref{eq:a12}. One can further show that the orthogonality of the row vectors of \eqref{eq:a12} implies that the second and third order-two subunitary matrices of the middle two rows of  \eqref{eq:a12} is related to each other by a linear relation. Hence, the center order-two submatrix of \eqref{eq:a12} decides its three adjacent order-two subunitary matrices in the lower right corner of \eqref{eq:a12}. In other word, \eqref{eq:a12} is the function of the parameters $z_1,z_2,z_3,z_4$ and $a_1,a_2$. It is known that $z_2,z_3,z_4$ are the functions of $z_1$ by \cite[Theorem 11]{karlsson11}. So \eqref{eq:a12} are the functions of three parameters $z_1,a_1$ and $a_2$. This fact coincides with \cite[Theorem 11]{karlsson11}.

Finally, the entangling power evaluates the maximum entanglement a bipartite unitary gate can create by acting on product states as input. It has been extensively investigated recently \cite{cy16}. We can show that the entangling power of the bipartite unitary gate $T_{AB}$ in the proof of Lemma~\ref{le:chm} (viii) is at least $1$ ebit, by using the input state ${1\over\sqrt2}(\ket{00}+\ket{11})_{AR_A}\ox{1\over\sqrt2}(\ket{00}+\ket{11})_{BR_B}$ with the reference systems $R_A$ and $R_B$.


\subsection{MUBs}
\label{subsec:mub}

In this subsection we investigate the MUBs. We say a vector is dephased if it is a zero vector or if its first nonzero element is real and positive. For any order-$d$ unitary $U$, we denote \textit{an MU vector} of $U$ as a dephased normalized vector unbiased to all column vectors of both $I_d$ and $U$. Let $N_v(U)$ denote the number of such vectors. Such vectors provide examples of the so-called zero noise, zero disturbance (ZNZD) states for two orthonormal bases (the standard basis and the basis represented by the columns of $U$) \cite{kjr14}.

We show that there is a vector unbiased to any two orthonormal bases in any dimension in Lemma \ref{le:mubsinkhorn}. We discuss $N_v(U)$ in Conjecture~\ref{cj:muvectors} and Lemma~\ref{lm:muvectors}. The main result of this subsection is Lemma \ref{le:mub}. We shall construct several order-six CHMs that cannot be a member of any MUB trio.

\bl\label{le:mubsinkhorn}
Let $d$ be an integer greater than $1$.
\\
(i) For any two orthonormal bases in $\bbC^d$, there is a normalized vector unbiased to both bases.
Equivalently, for any unitary matrix $U$ of order $d$, we have $N_v(U)\ge1$.
\\
(ii) For any two MUBs in $\bbC^d$, there is a normalized vector unbiased to both MUBs.
Equivalently, for any CHM $U$ of order $d$, we have $N_v(U)\ge1$.
\el
\bpf
(i)  Suppose $U$ is an order-$d$ unitary matrix. It follows from Lemma \ref{le:la} (vii) that there are diagonal unitaries $L$ and $R$ such that $B:=L U^\dag R$ is doubly quasistochastic (i.e. $B$ satisfies that its row and column sums are all one). Denote the column vectors of $U$ as $\vec{v}_i$, $i=1,2,\dots,d$, then the row vectors of $U^\dag$ are $\vec{v}_i^\dg$. Since the $B$ has its row sums being one, we have $B \vec{e}=\vec{e}$, where $\vec{e}$ is the vector with all its $d$ elements being $1$. Let the column vector formed by the diagonal elements of $R$ be $\vec{f}$. The equations $B=L U^\dag R$ and $B \vec{e}=\vec{e}$ imply that $L U^\dag R \vec{e}=\vec{e}$. Since $R \vec{e}=\vec{f}$, the previous equation can be written as $U^\dag \vec{f}=L^\dag \vec{e}$. Hence, the inner product of $\vec{v}_i$ and $\vec{f}$, i.e., $\sum_j [v_i(j)]^\ast f(j)$ is of unit modulus, where $v_i(j)$ and $f(j)$ are the elements of $\vec{v}_i$ and $\vec{f}$, respectively. Then, the vector $\vec{f}/\sqrt{d}$ is the desired normalized vector unbiased to all column vectors of $U$.

(ii) The assertion is a corollary of (i).
\epf

Lemma~\ref{le:mubsinkhorn} is a direct consequence of the Sinkhorn normal form of unitary matrices proved in \cite{iw15}. It is essentially known in \cite{kjr14}, as pointed out by \cite{idel16}. It also appears to be known in \cite{ab17}. The claim that there are at least $2^{d-1}$ unbiased vectors for any two given orthonormal bases in \cite{kjr14} has met some challenge in \cite{ab17}. i.e., when $d=3$, the number of unbiased vectors could be only $3$ for some pair of bases.

\bcj\label{cj:muvectors}
(i) For any CHM $U$ of order six, $N_v(U)$ is finite.
\\
(ii) $N_v(U)$ has a constant upper bound for all order-six CHM $U$.
\ecj
Although $(ii)\rightarrow (i)$ holds, the converse may be wrong. The reason is that when (i) holds, there may be an infinite sequence of order-six CHMs $U_j$, $j\in\mathbb{N}$, such that $\sup_{j\in\mathbb{N}} N_v(U_j)=\infty$, while for any $U_j$, $N_v(U)$ is finite.

Analogies of Conjecture \ref{cj:muvectors} are known to be true for dimensions $2,3$ and $5$, but false for dimension $4$ \cite{mub09}. The analogy is also false for any dimension $n$ divisible by a square, since Backelin \cite{Backelin89} proved that for such $n$, the number of ``cyclic $n$-roots'' is infinite. This is equivalent to that $N_v(F_n)$ is infinite, where $F_n$ is the Fourier matrix of order $n$. On the other hand according to \cite{mub09}, all the CHMs of order $2,3,5$ satisfy that $N_v(U)$ is a finite even number. Such statement for the classes of order-six CHMs considered in \cite{mub09} is supported by numerical evidence presented in \cite{mub09,Goyeneche13}. The classes of order-six CHMs considered in \cite{mub09} include the bicirculant Hadamard matrix, see also the third paragraph of \cite[p87]{deb10}. There are also some classes not considered in \cite{mub09}. There are some analytically proven cases in \cite{mub09}, such as the Fourier matrix $F_6$.

The following lemma is independent of Lemma~\ref{le:mubsinkhorn}, although its proof cites the proof of Lemma~\ref{le:mubsinkhorn}.

\bl\label{lm:muvectors}
Let $U$ be a unitary matrix of order $d$. Let $Q_1$ and $Q_2$ be arbitrary order-$d$ complex permutation matrices. Then\\
(i) $N_v(U)$, $N_v(U^\dag)$, $N_v(U^\ast)$, $N_v(U^T)$ are equal and they may be infinite.\\
(ii) $N_v(U)$ and $N_v(Q_1 U Q_2)$ are equal and they may be infinite.
\el
\bpf
Suppose the Sinkhorn normal form of $U$ is $U=D_1 B D_2$, where $D_1$ and $D_2$ are diagonal unitaries, and $B$ is doubly quasistochastic. We have $U^\dag=D_2^\dag B^\dag D_1^\dag$, where $B^\dag$ is also doubly quasistochastic. From the proof of Lemma~\ref{le:mubsinkhorn}, it can be figured out that a MU vector of $U$ is just the diagonal of $D_1/\sqrt{d}$ rewritten as a vector, thus it is in one-to-one correspondence with $D_1$. Similarly, a dephased normalized vector unbiased to $I_d$ and $U^\dag$ is in one-to-one correspondence with $D_2$. It is easy to see that the $D_1$ and $D_2$ in the Sinkhorn normal form of $U$ are in one-to-one correspondence with each other for a fixed $U$. Thus there is a bijection between the MU vectors of $U$ and those of $U^\dag$. Hence $N_v(U)$ and $N_v(U^\dag)$ are equal. They are both infinite when $U=I$. It can also be shown that they are both infinite whenever $U$ is block-diagonal, by multiplying the unbiased vectors for individual blocks by suitable factors and concatenating them into a normalized vector.

From $U^\ast=D_1^\ast B^\ast D_2^\ast$, we similarly have that the MU vectors of $U^\ast$ are in one-to-one correspondence with $D_1$. Thus  $N_v(U)$ and $N_v(U^\ast)$ are equal. The $N_v(U^T)$ is also equal to them by applying the assertion that $N_v(U)$ and $N_v(U^\dag)$ are equal to $U^\ast$. This completes the proof of (i).

(ii) The equality follows from the definition of $N_v(U)$. The possibility that they may be infinite is shown in the proof of (i).
This completes the proof.
\epf

\smallskip
In the following lemma, we first review a few known results on the product MUBs. Then we construct several order-six CHMs each of which cannot be a member of any MUB trio.

\bl
\label{le:mub}
(i) If a normalized vector is mutually unbiased to $d-1$ vectors in an orthonormal basis in $\bbC^d$, then it is also unbiased to the $d$'th vector in the basis.
\\
(ii) An order-six CHM is a member of some MUB trio if and only if so is its adjoint matrix, if and only if so is its complex conjugate, and if and only if so is its transpose.
\\
(ii.a) Let $k$ be a positive integer at most three. Then $k$ order-six CHMs are the members of some MUB trio if and only if so are their complex conjugate.
\\
(iii) Any set of three product-vector MUBs in the space $\bbC^2\ox\bbC^3$ is LU equivalent to either
\bea
\cT_0:=\{\ket{a_j,d_k},\ket{b_j,e_k},\ket{c_j,f_k}\}
\eea
or
\bea
\cT_1:=\{\ket{a_j,d_k},\ket{b_j,e_k},\ket{c_0,f_k},\ket{c_1,g_k}\},
\eea
where $\{\ket{a_j}\},\{\ket{b_j}\}$ and $\{\ket{c_j}\}$ is a complete set of MUBs in $\bbC^2$, and $\{\ket{d_j}\},\{\ket{e_j}\},\{\ket{f_j}\}$ and $\{\ket{g_j}\}$ is a complete set of MUBs in $\bbC^3$.
\\
(iv) Any set of four MUBs in $\bbC^6$ contains at most one product-vector basis. Equivalently, any two of four MUBs in $\bbC^6$ contain at most ten product vectors.
\\
(v) Any MUB trio contains none of the thirteen order-six CHMs
\bea
\label{eq:3x3subunitary}
&&
Y_1,
\\\label{eq:3x2}
&&
Y_2,
\\\label{eq:3x31ortho2}
&&
Y_3,
\\\label{eq:3prod}
&&
Y_4,
\\\label{eq:3x3singular}
&&
Y_5,
\\\label{eq:3x2real}
&&
Y_6,
\\\label{eq:2prod=ab,ac}
&&
Y_7,
\\\label{eq:adjoint}
&&
Y_8,
\\\label{eq:2x2subunitary+2x2subsingular}
&&
Y_9,
\\\label{eq:4x3}
&&
Y_{10},
\\\label{eq:3x3v}
&&
Y_{11},
\\
\label{eq:d1vd2}
&&
\bigg(
U\ox I_3\bigg)
\cdot
\left(
                   \begin{array}{cccccc}
                     V_1 &  0 \\
                     0 &  D_1V_1D_2 \\
                     \end{array}
                 \right)
\cdot
\bigg(
X \ox I_3\bigg),
\\
\label{eq:xi3}
&&
\left(
                   \begin{array}{cccccc}
                     V_2 &  0 \\
                     0 &  W_2 \\
                     \end{array}
                 \right)
\cdot
\left(
                   \begin{array}{cccccc}
                     D_3 &  D_4 \\
                     D_4 &  -D_3 \\
                     \end{array}
                 \right)
\cdot
\left(
                   \begin{array}{cccccc}
                     V_3 &  0 \\
                     0 &  W_3 \\
                     \end{array}
                 \right),
\eea
where
\bem
\item
$Y_1$ contains an order-three subunitary submatrix.
\item
$Y_2$ contains a submatrix of size $3\times2$ and rank one.
\item
$Y_3$ contains an order-three submatrix whose one column vector is orthogonal to the other two column vectors.
\item
three column vectors of $Y_4$ are product vectors.
\item
$Y_5$ contains an order-three singular submatrix.
\item
$Y_6$ contains a real submatrix of size $3\times2$.
\item
two column vectors of $Y_7$ are product vectors $\ket{a,b}$ and $\ket{a,c}$.
\item
$Y_8$ contains an order-four submatrix.
\bea
\label{eq:y8}
{1\over\sqrt6}
\left(
                   \begin{array}{cccccc}
                     1 &  1 & 1 & 1  \\
                     1 &  1 & x & y^* \\
                     1 &  x^* & 1 & z \\
                     1 &  y & z^* & 1 \\
                     \end{array}
                 \right),
\eea
where $x,y,z$ are complex numbers of modulus one.
\item
$Y_9$ contains an order-two subunitary matrix and an order-two singular matrix at the same time. The two matrices belong to exactly two columns of $Y_9$.
\item
$Y_{10}$ contains a submatrix of size $4\times3$ whose one column vector is orthogonal to the other two column vectors.
\item
The first three columns of $Y_{11}$ can be written as $\left(
                   \begin{array}{cccccc}
                     D_1 V D_2 \\
                     V \\
                     \end{array}
                 \right)$, where $D_1$ and $D_2=\left(
                   \begin{array}{cccccc}
                     0 & a & 0 \\
                     0 & 0 & b \\
                     c & 0 & 0 \\
                     \end{array}
                 \right)$ are both order-three GPMs.
\item
In Eq.~\eqref{eq:d1vd2}, $U$ and $X$ are both order-two unitary matrices, $V_1$ is an order-three unitary matrix, and $D_1,D_2$ are both order-three diagonal unitary matrices.
\item
In Eq.~\eqref{eq:xi3}, $V_2,V_3,W_2$ and $W_3$ are all order-three unitary matrices, $D_3$ and $D_4$ are both diagonal matrices of real and nonnegative elements and (1) $V_3$ and $W_3$ both have four zero elements, or (2) one of $D_3$ and $D_4$ is singular, or (3) $D_3$ and $D_4$ are proportional, or (4) $V_3$ and $W_3$ respectively have one and six zero elements.
\eem
\qed
\el
We prove Lemma \ref{le:mub} in Appendix \ref{app:{le:mub}}. We believe that statement (ii.a) does not hold when the ``complex conjugate'' is replaced by the ``transpose''. However we do not have a counterexample yet. If the statement with the replacement was true, then the statement with another replacement would be also true by (ii.a), i.e., replace the ``complex conjugate'' by ``adjoint matrices''.

In Lemma~\ref{le:mub} (v), we note that $Y_1$ is a special case of $Y_3$, and $Y_2$ is a special case of $Y_6$ up to the equivalence of MUB trios.
The so-called Fourier family of Hadamard matrices $F(a,b)$ constructed in Eq. (3) of \cite{jmm09} [same as the family $F_6^{(2)}$ with its standard form shown in Eq.~\eqref{eq:ufourier1}] has a $2\times3$ submatrix of rank one. The main result of the paper \cite{jmm09} says that $F(a,b)$ cannot be a member of any MUB trio.
Since the matrix $Y_2$ contains $F(a,b)$, the main conclusion of \cite{jmm09} is included as a special case of Lemma \ref{le:mub}. However, it should be noted that Lemma \ref{le:mub} depends on \cite{jmm09}, see the end of Sec.~\ref{subsec:mubtrio} for details. Further, the matrix $Y_8$ is the same as the matrix $H$ in \cite[Lemma 2.7.]{nicoara08} and plays the fundamental role in studying the self-adjoint order-six CHMs \cite{nicoara08}.

Further, Lemma~\ref{le:mub} (v) shows the fact that a CHM cannot be a member of any MUB trio when the CHM contains a submatrix of size $k\times 3$ and one column vector of the submatrix is orthogonal to the other two column vectors. Indeed the lemma has proved the fact when $k=2,3,4$. One can straightforwardly show that $k\ne1$ and $5$, because the CHM has only nonzero elements.

As mentioned in the Introduction, Theorem~\ref{thm:2prod} is a direct consequence of the case of $Y_4$ in Lemma~\ref{le:mub} (v). We apply Theorem~\ref{thm:2prod} and Lemma \ref{le:mub} to investigate the entanglement property of the column states in an MUB trio.

\bl
\label{le:entanglement}
Suppose $\ket{\a}$, $\ket{\b}$, and $\ket{\g}$ are three normalized states from three column vectors of the same basis in an MUB trio, and $\ket{\a}$, $\ket{\b}$ are product states. If $p\ge0,q\ge0$ and $p+q<1$, then
the quantum state $p\proj{\a}+q\proj{\b}+(1-p-q)\proj{\g}$ is entangled and distillable under LOCC.
\el
\bpf
It follows from the matrix $Y_7$ and $Y_2$ in Lemma \ref{le:mub} (v) that $\ket{\a}=\ket{a,b}$, $\ket{\b}=\ket{a',c}$, where $\ket{a}$ and $\ket{a'}$ are linearly independent normalized states in $\bbC^2$, while $\ket{b}$ and $\ket{c}$ are linearly independent normalized states in $\bbC^3$. It follows from Theorem \ref{thm:2prod} that $\ket{\g}$ is entangled. Let $\ket{\g}=\ket{0,d}+\ket{1,e}$ where $\ket{d}$ and $\ket{e}$ are linearly independent unnormalized states. It follows from the matrix $Y_5$ in Lemma \ref{le:mub} (v) that $\ket{d},\ket{e}\not\in\lin\{\ket{b},\ket{c}\}$. Let $U\ox V$ be an invertible order-six matrix such that $U\ket{a}=\ket{0}$, $U\ket{a'}=\ket{1}$ and $(U\ox V)\ket{\g}=\frac{1}{\sqrt{2}}(\ket{00}+\ket{11})$. Hence $V\ket{d},V\ket{e}\in\bbC^2$. Since $\ket{b}$ and $\ket{c}$ are linearly independent, one of $V\ket{b}$ and $V\ket{c}$ is not in $\bbC^2$. We have
\bea
\r
&:=&(U\ox V)
(p\proj{\a}+q\proj{\b}+(1-p-q)\proj{\g})
(U^\dg \ox V^\dg)
\notag\\
&=&
p\proj{0}\ox V\proj{b}V^\dg
+
q\proj{1}\ox V\proj{c}V^\dg
+
\frac{1}{2}(1-p-q)(\ket{00}+\ket{11})(\bra{00}+\bra{11}).
\eea
Since one of $V\ket{b}$ and $V\ket{c}$ is not in $\bbC^2$, we may assume it as $\ket{2}$ by performing an invertible basis transform $I_2\ox W$ on $\r$ such that $W\ket{0}=\ket{0}$ and $W\ket{1}=\ket{1}$. One can verify that the partial transpose of the $2\times3$ state $(I_2 \ox W) \r (I_2 \ox W^\dg)$ is not positive semidefinite. Hence $\r$ is entangled~\cite{hhh96}, and so is $p\proj{\a}+q\proj{\b}+(1-p-q)\proj{\g}$.

Since $\r$ is an $2\times 3$ entangled state, the Peres-Horodecki criterion~\cite{hhh96} implies that $\r$ is not PPT. Such states are distillable under LOCC~\cite{hhh97}. 
This completes the proof.
\epf

\subsection{The Schmidt rank of matrices in an MUB trio}
\label{subsec:mubtrio}

In this subsection we introduce the main results of this paper. We characterize the CHMs of Schmidt rank one, two and  three by giving equalities and inequalities involving their parameters in Lemma \ref{le:sr2}. Then we show that the member of any MUB trio has Schmidt rank at least three in Theorem \ref{thm:sr3}. We further show in Corollary \ref{cr:sr3} that
if a set of four MUBs in $\bbC^6$ contains the identity, then any other MUB in the set has min-Schmidt rank at least three.

\bl
\label{le:sr2}
(i) Any Schmidt-rank-one order-six CHM is locally equivalent to
\bea
\bbH_1:=
{1\over\sqrt6}
\left(
                   \begin{array}{cccccc}
                     1 &  1 \\
                     1 &  -1 \\
                     \end{array}
                 \right)
\ox
\left(
                   \begin{array}{cccccc}
                     1 &  1 & 1 \\
                     1 &  \o & \o^2 \\
                     1 &  \o^2 & \o \\
                     \end{array}
                 \right).
\eea

(ii) Up to a product complex permutation matrix from the right hand side, any Schmidt-rank-two order-six CHM can be written as
\bea
\label{eq:v=w}
\bbH_2(\a,\b,\g,V,W):=
\bigg[
\left(
                   \begin{array}{cccccc}
                     \cos\a &  \sin\a \\
                     e^{i\g}\sin\a &  -e^{i\g}\cos\a \\
                     \end{array}
                 \right)
\ox I_3\bigg]
\cdot
\left(
                   \begin{array}{cccccc}
                     V &  0 \\
                     0 &  W \\
                     \end{array}
                 \right)
\cdot
\bigg[
\left(
                   \begin{array}{cccccc}
                     \cos\b &  \sin\b \\
                     \sin\b &  -\cos\b \\
                     \end{array}
                 \right)
\ox I_3\bigg],
\eea
where $V=[v_{jk}]$ and $W=[w_{jk}]$ are linearly independent
order-three unitary matrices and
\bea
&&
\label{eq:sr2-0}
\a,\b\in[0,\p/4],~~~~\a+\b\ge\p/4,~~~~\g\in[0,2\p),
\\&&
\label{eq:sr2-1}
\cos2\a\cos2\b+{3(v_{jk}w_{jk}^*+v_{jk}^*w_{jk})\over2}\sin2\a\sin2\b=0,
\\&&\label{eq:sr2-2}
\abs{v_{jk}}^2+\abs{w_{jk}}^2=2/3,
\\&&\label{eq:sr2-3}
(\abs{v_{jk}}^2-1/3)\cos2\a=0,
\\&&\label{eq:sr2-4}
(\abs{v_{jk}}^2-1/3)\cos2\b=0.
\eea
Hence
\\ (ii.a) If $(\a,\b)\ne(\p/4,\p/4)$ then  \eqref{eq:sr2-0} and \eqref{eq:sr2-1} both hold, $V$ and $W$ are both CHMs.
\\ (ii.b) If $(\a,\b)=(\p/4,\p/4)$ then $\g\in[0,2\p)$, $v_{jk}^*w_{jk}+v_{jk}w_{jk}^*=0$, and $\abs{v_{jk}}^2+\abs{w_{jk}}^2=2/3$.
\\ (ii.c) If $\bbH_2(\a,\b,\g,V,W)$ is a member of some MUB trio, then $\a+\b>\p/4$.
\\ (ii.d) $\bbH_2(\a,\b,\g,V,W)$ is not a member of any MUB trio.

(iii) Any Schmidt-rank-three order-six CHM can be written as
\bea
\label{eq:sr3}
&&
\bbH_3(\a_1,\b_1,\g_1,\a_2,\b_2,\g_2,\a_3,\b_3,\g_3,V,W):=
\\&&
(I_2 \ox V)\cdot
\left(
                   \begin{array}{cccccc}
                     \cos\a_1 &  0 & 0 & e^{i\g_1}\sin\a_1 & 0 & 0 \\
                     0 &  \cos\a_2 &  0 & 0 & e^{i\g_2}\sin\a_2 & 0 \\
                     0 & 0 &  \cos\a_3 &  0 & 0 & e^{i\g_3}\sin\a_3 \\
                    e^{i\b_1}\sin\a_1 &  0 & 0 & -e^{i(\b_1+\g_1)}\cos\a_1 & 0 & 0 \\
                     0 &  e^{i\b_2}\sin\a_2 &  0 & 0 & -e^{i(\b_2+\g_2)}\cos\a_2 & 0 \\
                     0 & 0 &  e^{i\b_3}\sin\a_3 &  0 & 0 & -e^{i(\b_3+\g_3)}\cos\a_3 \\
                     \end{array}
                 \right)
\cdot
(I_2 \ox W),
\notag
\eea
where $V$ and $W$ are
order-three unitary matrices, the first column vector of $W$ have all nonnegative and real elements, and
\bea
&&
\label{eq:sr3,abgde}
\a_1,\a_2,\a_3\in[0,\p/2],~~~~\b_1,\b_2,\b_3,\g_1,\g_2,\g_3\in[0,2\p).
\eea
Hence
\\ (iii.a) If $\bbH_3(\a_1,\b_1,\g_1,\a_2,\b_2,\g_2,\a_3,\b_3,\g_3,V,W)$ is a member of some MUB trio, then $\a_1,\a_2,\a_3\in(0,\p/2)$.
\qed
\el
We prove Lemma \ref{le:sr2} in Appendix \ref{app:{le:sr2}}. For more examples of equivalent matrices we refer the readers to \cite{mub09}.
We point out that the unitary matrices $V$ and $W$ in case (ii.b) may be not CHMs. Besides, the order-six CHM in (ii.b) can be written as
$\left(
                   \begin{array}{cccccc}
                    S & T \\
                     T & S \\
                     \end{array}
                 \right)$ where $S$ and $T$ are order-three matrices and may be not proportional to unitary matrices. An example for the above two statements is the so-called Björck's circulant matrix, see \cite[Eq. (A2)]{bw09}.
On the other hand,
Eqs. \eqref{eq:sr2-0}-\eqref{eq:sr2-4} guarantee that the matrix in \eqref{eq:v=w} is an order-six CHM of Schmidt rank two.
Now we are in a position to prove Theorem \ref{thm:sr3}.
\bpf
Any MUB trio contains no matrix of Schmidt rank one because of Lemma \ref{le:mub} (iv) and Lemma \ref{le:sr2} (i). Any MUB trio contains no matrix of Schmidt rank two because of Lemma \ref{le:sr2} (ii) and (ii.d). This completes the proof.
\epf

From Theorem \ref{thm:sr3} and Definition \ref{df:mub}, we obtain
\bcr
\label{cr:sr3}
If a set of four MUBs in $\bbC^6$ contains the identity, then any other MUB in the set has min-Schmidt rank at least three.
\ecr

From Theorem \ref{thm:sr3} and Lemma \ref{le:chm} (ix), we obtain
\bcr
\label{cr:fourierfamily}
Any CHM in the Fourier family $F_6^{(2)}$ or the Fourier transposed family $(F_6^{(2)})^T$ is not a member of any MUB trio.
\ecr
The first part of this statement has been proven in \cite{jmm09}, and also by the case of the matrix $Y_2$ in Lemma \ref{le:mub} (v). The second part on the Fourier transposed family follows from the first part because of Lemma \ref{le:mub} (ii). The two proofs of Corollary \ref{cr:fourierfamily} are essentially computer-aided proofs, since they rely on \cite{mw12jpa102001} which uses the result on the Fourier family in \cite{jmm09}, and the proof in the latter paper is computer-aided. Further, they are currently not independent from \cite{jmm09}. However, it is possible that the connection with product vectors or the Schmidt rank introduced by the two proofs in this paper may lead to an independent proof for Corollary \ref{cr:fourierfamily}.

\subsection{Numerical Studies}
\label{ssec:num}

By using a numerical search method proposed in \cite{Goyeneche13}, we have found 90 normalized and dephased vectors that are approximately unbiased to the column vectors of both the identity matrix $I_6$ and the spectral matrix in Eq.~\eqref{eq:spec}. This agrees with the results in \cite{bw09,Goyeneche13}. The result of 90 exactly unbiased vectors is a rigorous result of \cite{bw09}. The method of \cite{Goyeneche13} alone cannot rigorously imply the same result. The same numerical method is also presented in \cite{fr15}. We calculated the absolute value of the inner product of all pairs of the 90 approximate vectors, and it is always greater than $0.1$. Thus we obtain the following statement.

Assume that the following two conditions hold: (i) The spectral matrix $S$ in Eq.~\eqref{eq:spec} satisfies that $N_v(S)=90$; (ii) The method in \cite{Goyeneche13}, when applied to the spectral matrix, gives solutions that approximates the exact solutions to an accuracy defined as $\vert\braket{\phi}{\psi}\vert<0.01$, where $\ket{\phi}$ and $\ket{\psi}$ are the exact state and the approximate state, respectively. Under these two assumptions, no two of the 90 exact vectors are orthogonal to each other. In other words, there are no two normalized and mutually orthogonal vectors in $\bbC^6$ that are unbiased to the column vectors of both $I_6$ and the spectral matrix.

We call the last sentence ``the assertion.'' It is stated in \cite{bw10}, which uses semidefinite programming to reach the conclusion. The assertion is also confirmed by direct calculations on the $90$ exact vectors provided to us by M. Grassl, who obtained them by exactly solving some polynomial equations, see the method in \cite{Grassl04} and also \cite{bw09}.

The method of numerical search in \cite{Goyeneche13} is not proven to be rigorous. It may be possible to make the assertion above unconditional by extending the numerical effort in \cite{Goyeneche13}, but some difficulty is as follows: when the inner product of two vectors is near zero, the deviation from unbiased status of both vectors (with respect to a given order-six CHM, which is the spectral matrix in the current case) may be very small. On the other hand, the method of \cite{bw09} escapes this difficulty by solving  some polynomial equations. Such phenomenon of small deviation from the unbiased status has occurred in \cite{rle11}, which tries to find four almost mutually unbiased bases.

The method of numerical search in \cite{Goyeneche13} shares some similarity with that for finding the Sinkhorn normal form of a unitary matrix in \cite{vb14}, but they are still quite different. In fact the two methods can be viewed as dual to each other in some sense: the former performs iterations on the vector while the latter does so on the matrix.

Numerical evidence suggests that the Dita matrix $D_0$ (\cite[Eq. (A1)]{bw09}) may be of min-Schmidt rank two. The $D_0$ multiplied by phase $i$ on one row contains a real $3\times 2$ submatrix, hence from the case $Y_6$ in Lemma~\ref{le:mub} (v), it cannot appear in a MUB trio. The $D_0$ is a special case of a one-parameter family called the Hermitean family in \cite{mub09}. This family is first introduced in \cite{nicoara08}. Two other special matrices in this family are $H(\gamma_0)$ and $H(-\gamma_0)$ in the notation of \cite{nicoara08}, where $\gamma_0=\arccos^{-1}(\frac{\sqrt{3}-1}{2})$ (note that $\gamma_0$ is different from the $\theta_0$ in \cite{mub09}). They are complex conjugate of each other, so they have the same min-Schmidt rank according to Lemma~\ref{le:la} (x). Numerical evidence suggests that they may be of min-Schmidt rank two. Numerical evidence also suggests that $H(\gamma_0)$ and $H(-\gamma_0)$ multiplied by randomly chosen diagonal unitaries on the two sides are of Schmidt rank four, and that all matrices in the Hermitean family are of min-Schmidt rank at most three.

Numerical evidence suggests that there are many $H_2$-reducible CHMs that do not contain an order-three subunitary submatrix. We have tested the random instances among the two special classes of CHMs in \cite[Sec. 6]{karlsson11}, and found that they always do not contain an order-three subunitary submatrix. There is at least one counterexample, e.g. the case $z=1$ in the latter class. The reason this is of interest is that if an order-six CHM contains an order-three subunitary submatrix, then it cannot appear in any MUB trio in terms of $Y_1$ in Lemma~\ref{le:mub} (v).

\section{Open problems}
\label{sec:pro}
The main results of this paper are Theorem \ref{thm:2prod} and Theorem \ref{thm:sr3}. Some applications to specific partial sets of MUBs have been presented in various sections including Sec.~\ref{ssec:num}. Lemma \ref{le:entanglement} exemplifies a potentially interesting link to the topic of mixed state entanglement as a physical application. Two main problems arise from this paper. First, what if we replace the identity matrix in Theorem \ref{thm:2prod} with some other matrix? Can we still obtain the same conclusion? Second, can we show that no Schmidt-rank-three CHM can be a member of some MUB trio? It would improve Theorem \ref{thm:sr3}. Investigating the two problems would further improve our understanding towards the existence of four MUBs in $\bbC^6$. We shall also study Conjecture~\ref{cj:muvectors}. Related to the second question above, a possible topic for future study is the min-Schmidt rank of CHMs, which requires studying the CHMs up to equivalence.
\smallskip

\section*{Acknowledgments}

We thank Ingemar Bengtsson for comments about the relationship between the Sinkhorn normal form of unitary matrices and the existence of vectors unbiased to two given orthonormal bases. We thank Markus Grassl for helpful comments and for sending us the exact expressions for the 90 vectors unbiased to the columns of both the order-six identity matrix and the spectral matrix. L.C. was supported by Beijing Natural Science Foundation (4173076), the NNSF of China (Grant No. 11501024), and the Fundamental Research Funds for the Central Universities (Grant Nos. KG12001101, ZG216S1760 and ZG226S17J6). L.Y. acknowledges support from the Ministry of Science and Technology of China under Grant No. 2016YFA0301802, the funds of Hangzhou City for supporting the Hangzhou-City Quantum Information and Quantum Optics Innovation Research Team, a grant from the Department of Education of Zhejiang Province, the startup grant of Hangzhou Normal University, and partial support by NICT-A (Japan).

\bibliography{channelcontrol}

\appendix

\section{The proof of Lemma~\ref{le:la}}

\label{app:{le:la}}

\bpf
(i) can be proved straightforwardly.

(ii) Since $[u_{jk}]$ and $[p_{jk}u_{jk}]$ are both unitary we have
\bea
\label{eq:sum3}
\sum^3_{j=1} u_{1j}^* u_{2j} = \sum^3_{j=1} p_{2j} u_{1j}^* u_{2j} = 0.
\eea
Since $p_{21}=p_{31}=1$, if $[u_{jk}]$ contains no zero then $p_{jk}=1$ for all $j,k$. So the assertion holds. Suppose $[u_{jk}]$ contains exactly one zero entry.
Eq. \eqref{eq:sum3} implies that $p_{22}=p_{23}$. We can similarly prove that
$p_{32}=p_{33}$, and construct $D_1$ and $D_2$.

(iii) Suppose $\{\ket{\ps_j}\}_{j=1,\cdots,6}$ is the basis and $\ket{\ps_j}=\ket{a_j,b_j}$  for $j=1,\cdots,5$. Since $\sum^6_{j=1} \proj{\ps_j}=I_6$, we have
\bea
\proj{\ps_6}=I_6-\sum^5_{j=1} \proj{a_j,b_j}.
\eea
We perform the partial transpose on both sides of the above equation, and both sides are still positive semidefinite matrices. It holds only if $\ket{\ps_6}$ is a product state.

(iv) Since $e^{ia}+e^{ib}=e^{ic}+e^{id}$, we have $\cos a+\cos b=\cos c+\cos d$ and $\sin a+\sin b=\sin c+\sin d$. By squaring both sides and summing up them, we obtain
\bea
\cos(a-b)=\cos(c-d),
\\
\cos(a-c)=\cos(b-d),
\\
\cos(a-d)=\cos(b-c).
\eea
Straightforwardly solving the equations lead to the assertion.

(v) It suffices to prove that claim that if $V_0$ is singular then so is $V_3$.
Suppose $V_0$ and $V_3$ are respectively of size $m\times n$ and $(d-m)\times (d-n)$. Suppose $V_0$ has rank $r<n$.
Let
$W$ be an order-$m$ unitary matrix such that the first $m-r$ row vectors of $WV_0$ is zero.
Since $(W\op I_{d-m}) U=\left(
                   \begin{array}{cccccc}
                     WV_0 & WV_1 \\
                     V_2 & V_3 \\
                     \end{array}
                 \right)$ is unitary, the first $m-r$ row vectors of $(W\op I_{d-m}) U$ are orthogonal to the row vectors of $\left(
                   \begin{array}{cccccc}
                     V_2 & V_3 \\
                     \end{array}
                 \right)$. Hence the first $m-r$ row vectors of $WV_1$, which are pairwise orthogonal vectors, are orthogonal to the row vectors of $V_3$. Since $(d-m)+(m-r)>d-n$, we obtain that the row vectors of $V_3$ are linearly dependent. So $V_3$ is singular.

To prove the second claim, we assume that $V_0$ has order $m$ and rank $r$. Then
\bea
\rank V_0 + \rank V_3 \le r + (d-m)-(m-r)=d-2m+2r=d-2\dim\ker V_0.
\eea

(vi) The assertion can be proved straightforwardly.

(vii) The assertion follows from the Sinkhorn normal form of unitary matrices \cite[Theorem 5.13]{idel16}, and the proof for that theorem is in \cite{iw15}.

(viii) The assertion follows from the Sinkhorn normal form of positive definite matrices \cite[Theorem 5.12]{idel16}.

(ix) Let $\ket{a_1,b_1},\cdots,\ket{a_k,b_k}$ be the $k$ product states of the orthonormal basis in the hypothesis. Let $\ket{c_{k+1}},\cdots,\ket{c_6}$ be the remaining vectors in the basis. Hence $\r:=\sum^6_{j=k+1} \proj{c_j} = I_6 - \sum^k_{j=1} \proj{a_i,b_i}$. Since $\ket{a_j,b_j}$ are orthonormal, $\r$ is a non-normalized positive partial transpose (PPT) state. The Horodecki-Peres criterion implies that $\r$ is a non-normalized $2\times3$ separable state and $\rank\r=\rank\r^\Gamma=6-k$. It follows from \cite[Table II]{cd12} that $\r=\sum^6_{j=k+1} \proj{e_j,f_j}$ with some product vectors $\ket{e_j,f_j}$. Since $\r$ is proportional to a projector, $\ket{e_j,f_j}$ are pairwise orthogonal.

(x) The assertion follows from the symmetry with respect to the reflection about the diagonal line, and the symmetry under complex conjugation.

This completes the proof. \epf

\section{The proof of Lemma~\ref{le:chm}}

\label{app:{le:chm}}

\bpf
(i) follows from Lemma \ref{le:la} (i).

(ii) Up to the equivalence, we may assume that the upper left submatrix of size $2\times k$ of the order-six CHM $U=[a_{ij}]_{i,j=1,\cdots,6}$ has rank one. Since
$
\sum^k_{i=1} a_{i1} a_{i2}^* = - \sum^6_{i=k+1} a_{i1} a_{i2}^*
$ and $\abs{a_{ij}}=1/\sqrt6$, we have
\bea
\abs{\sum^k_{i=1} a_{i1} a_{i2}^*} = k/6 = \abs{\sum^6_{i=k+1} a_{i1} a_{i2}^*} \le (6-k)/6.
\eea
So we have $k\le3$. The equality is achievable e.g., when $U={1\over\sqrt2}\left(
                   \begin{array}{cccccc}
                     1 &  1 \\
                     1 &  -1  \\
                     \end{array}
                 \right)\ox {1\over\sqrt3}\left(
                   \begin{array}{cccccc}
                     1 &  1 & 1 \\
                     1 &  \o & \o^2 \\
                     1 &  \o^2 & \o \\
                     \end{array}
                 \right)$.

(iii) Suppose some order-six CHM $U$ contains an order-three submatrix of rank one. We can find two complex permutation matrices $P$ and $Q$ such that the upper-left order-three submatrix $X$ of $PUQ$ have all elements $1/\sqrt6$, and the first column vector of $PUQ$ also have all elements $1/\sqrt6$. Since $PUQ$ is unitary, the second and third column vectors of the order-three submatrix of $PUQ$ below
$X$ are both equal to $-{1\over\sqrt6}(1,1,1)^T$. Then the second and third column vectors of $PUQ$ are not orthogonal. This is a contradiction with the unitarity of $PUQ$.

The second claim follows from Lemma \ref{le:la} (v).

Third, if $U$ contains a singular matrix of order five, then
it follows from Lemma \ref{le:la} (v) that $U$ contains a zero entry.
It is a contradiction with the fact that every entry of $U$ is nonzero.

(iv) The ``if'' part can be proved straightforwardly. Let us prove the ``only if'' part. Up to the equivalence, we may assume that the left-upper order-two submatrix of the CHM $U=[a_{ij}]$ is the subunitary matrix  ${1\over\sqrt6}\left(
                   \begin{array}{cccccc}
                     1 &  1 \\
                     1 &  -1  \\
                     \end{array}
                 \right)$, and $a_{j1}=1/\sqrt6$ for $j=3,4,5,6$. Since $U$ is unitary, we have $\sum^6_{j=3} a_{j2}=0$. Since $\abs{a_{j2}}=1/\sqrt6$, we have $a_{32}=-a_{42}$ and $a_{52}=-a_{62}$ up to a permutation of the subscripts by Lemma \ref{le:la} (iv). Up to the equivalence, the CHM contains two product column vectors $(1,1)^T/\sqrt2 \ox (1,1,1)^T/\sqrt3$ and $(1,-1)^T/\sqrt2 \ox (1,\sqrt6 a_{32},\sqrt6 a_{52})^T/\sqrt3$. By choosing $\ket{a},\ket{a^\perp}=(1,\pm1)^T/\sqrt2$ we have found the two product column vectors.

(v) The ``if'' part is trivial. We prove the ``only if'' part. Suppose the order-six CHM $U$ has a submatrix of size $3\times2$ with orthogonal column vectors. Up to equivalence we may assume that the submatrix is in the upper left corner of $U$, and it is ${1\over\sqrt6}\left(
                   \begin{array}{cccccc}
                     1 &  1 \\
                     1 &  \o  \\
                     1 &  \o^2  \\
                     \end{array}
                 \right)$. Up to equivalence we may assume that the lower left submatrix of size $3\times2$ of $U$ is ${1\over\sqrt6}\left(
                   \begin{array}{cccccc}
                     1 &  x \\
                     1 &  x\o  \\
                     1 &  x\o^2  \\
                     \end{array}
                 \right)$ with some complex number $x$ of modulus one. So the first two column vectors of $U$ are both product column vectors $\ket{a,b},\ket{c,b^\perp}$ where $\ket{b},\ket{b^\perp}$ is an orthonormal basis in $\bbC^3$.

(vi) The first assertion follows from the fact \cite[Theorem 11]{karlsson11}. The same fact implies that
\bea\label{eq:zia123}
\left(
                   \begin{array}{cccccc}
                     1 & 1 \\
                     z_3^* & -z_3^*\\
                     \end{array}
                 \right)\cdot \left(
                   \begin{array}{cccccc}
                     a_1 & a_2 \\
                     a_1a_3 & -a_2a_3\\
                     \end{array}
                 \right)\cdot \left(
                   \begin{array}{cccccc}
                     1 & z_1^* \\
                     1 & -z_1^*\\
                     \end{array}
                 \right)=\left(
                   \begin{array}{cccccc}
                     x & y \\
                     y^* & -x^*\\
                     \end{array}
                 \right),
\eea
where $x,y$ are complex numbers. By respectively comparing the diagonal elements and off-diagonal elements of the matrix in Eq.~\ref{eq:zia123}, we have
\bea
(a_1^* a_2^* a_3^*-z_1^* z_3^*)(a_1-a_1a_3-a_2-a_2a_3)&=&0,
\\
(a_1^* a_2^* a_3^*-z_1^* z_3^*)(a_1-a_1a_3+a_2+a_2a_3)&=&0.
\eea
Note that in deriving the first equation above we have used that $a_1^* a_2^* a_3^* (a_1-a_1a_3-a_2-a_2a_3)=a_2^* a_3^*-a_2^*-a_1^* a_3^*-a_1^*$.
If $a_1^* a_2^* a_3^*-z_1^* z_3^*\ne0$ then $a_1-a_1a_3=a_2+a_2a_3=0$. The equations have no solution. Hence $a_1^*a_2^*a_3^*-z_1^*z_3^*=0$ and
we obtain \eqref{eq:z1z3}. We apply the above argument to the remaining three order-two sub-unitaries of the lower right corner of \eqref{eq:a12} and obtain \eqref{eq:z2z3}-\eqref{eq:z2z4}, respectively. Next, \cite[Theorem 11]{karlsson11} implies that \eqref{eq:zia123} is equal to $\left(
                   \begin{array}{cccccc}
                     1 & 1 \\
                     z_4^* & -z_4^*\\
                     \end{array}
                 \right)\cdot \left(
                   \begin{array}{cccccc}
                     d_1 & d_2 \\
                     d_1d_3 & -d_2d_3\\
                     \end{array}
                 \right)\cdot \left(
                   \begin{array}{cccccc}
                     1 & z_2^* \\
                     1 & -z_2^*\\
                     \end{array}
                 \right)$. The equation, \eqref{eq:z1z3} and \eqref{eq:z2z4} imply \eqref{eq:a1+a3} and \eqref{eq:z3a1+a3}. One can similarly prove  \eqref{eq:b1+b3} and \eqref{eq:z3b1+b3}.

(vi.a) and (vi.b) are clear.


(vii) We prove for the order-four case and one can similarly prove the order-five case. If an order-six CHM $U$ contains an order-four subunitary submatrix, then there exist two permutation matrices $P$ and $Q$ such that the upper-left order-four submatrix $R$ of $PUQ$ is a subunitary matrix. Since $PUQ$ is an CHM, the four length-2 column vectors below $R$ are pairwise orthogonal. It is a contradiction with the fact that the vectors are all nonzero.

(viii) Left-multiply the spectral matrix $S$ in \eqref{eq:spec} by a diagonal matrix $\diag(1,\o,\o^2,\o,\o^2,\o)$, and right-multiply by the matrix
$
\left(
                   \begin{array}{cccccc}
                     0 & 1 & 0 & 0 & 0 & 0 \\
                     0 & 0 & 0 & 0 & 0 & \o \\
                     0 & 0 & 0 & 0 & 1 & 0 \\
                     0 & 0 & 1 & 0 & 0 & 0 \\
                     0 & 0 & 0 & \o & 0 & 0 \\
                     1 & 0 & 0 & 0 & 0 & 0 \\
                     \end{array}
                 \right),
$ we obtain $S'$ in \eqref{eq:spec2}. It has Schmidt rank three. Its upper left $3\times 3$ submatrix is a subunitary matrix, while the upper left and bottom right $3\times 3$ submatrices are equal. Hence the min-Schmidt rank of $S$ is at most three. Since $S'$ is of Schmidt rank three, it is equivalent to a controlled unitary under local unitaries \cite{cy14}. Using Eq.~\eqref{eq:spec2}, there are order-two diagonal unitary matrices $P_1$ and $P_2$, order-three unitary matrices $R_1$ and $R_2$ such that $S'=(P_1\ox R_1) T (P_2\ox R_2)$, and $T=\sum_{j=1}^3 A_j \ox \ketbra{j}{j}$, with $A_1={1\over \sqrt2}\left(
\begin{array}{cc}
1 & -1\\
1 & 1 \\
\end{array}\right)$,
$A_2={1\over \sqrt2}\left(
\begin{array}{cc}
1 & 1\\
-1 & 1 \\
\end{array}\right)$, and
$A_3={1\over \sqrt2}\left(
\begin{array}{cc}
1 & i\\
i & 1 \\
\end{array}\right)$.
In the following we prove that $S$ cannot be of min-Schmidt rank one or two.

Suppose $S$ has min-Schmidt rank one. There are complex permutation matrices $Q_1$ and $Q_2$ such that $T=Q_1 S Q_2$ is of Schmidt rank one. Suppose $T=\left(
                   \begin{array}{cc}
                     A & B \\
                     C & D \\
                     \end{array}
                 \right)$, where $A,B,C,D$ are $3\times 3$ matrices. Then these four blocks are proportional to each other. Suppose that under the action of $Q_2$, the first column of $S$ is mapped to the $k$-th column of $T$, where $1\le k\le 6$. Without loss of generality, we may assume that the first three elements of the $k$-th column of $T$ are all $1$ (since we may left-multiply $T$ by a diagonal matrix $I_2\ox D$ while preserving the Schmidt rank). Let $k'=k+3$ if $k\le 3$, and otherwise let $k'=k-3$. This means that the $k'$-th column of $T$ has all top three elements being the same. This is impossible, since by assumption the first three elements of the $k'$-th column of $T$ is the corresponding part of one of the later five columns of $S$ multiplied by a common phase factor, and those columns of $S$ do not have three identical elements in them. This proves that $S$ cannot be of min-Schmidt rank one.

Next, suppose $S$ has min-Schmidt rank two. There are complex permutation matrices $Q_1$ and $Q_2$ such that $T=Q_1 S Q_2$ is of Schmidt rank two. Suppose $T=\left(
                   \begin{array}{cc}
                     A & B \\
                     C & D \\
                     \end{array}
                 \right)$, where $A,B,C,D$ are $3\times 3$ matrices, and they may be called ``blocks'' below. Then there are exactly two linearly independent ones among the four blocks. Note that all elements of $T$ are of the same modulus $1/\sqrt{6}$. Let us view each block as a vector of length $9$ consisting of its elements. By applying Lemma~\ref{le:van} to these four vectors of length $9$, we obtain that either (a) there are at most two types of elements in the vectors of length $9$, such that the elements in each class are equal in each vector, or (b) each of the four blocks $A,B,C,D$ is proportional to one of the two among them.

Suppose the case (a) is true. At least five elements of the same positions in $A,B,C,D$ are the same in each block. There are two subcases. The case (a.1) is that there are three of these five elements that are in one row or column. In this case, without loss of generality we may assume the three elements are in the same row and that row is the first row in the blocks. Thus the inner product of the corresponding elements of the first row and the fourth row of $T$ must be either $1$ or $-1$, for the two rows to be orthogonal. But this is impossible, since when the $S$ is dephased with respect to any row (i.e. multiplying each column by a phase such that the given row becomes the all-one vector), all rows except the dephased row have three distinct elements $1,\o,\o^2$, each appearing twice. This excludes case (a.1). The case (a.2) is that no three of these five elements are in one row or column. Without loss of generality we may assume the five elements in the block $A$ are all $1$. The arrangement for the five elements is of one of the following two types, up to row and column permutations (The symbol ``$*$'' denotes other elements which may or may not be $1$):
                $\left(
                   \begin{array}{ccc}
                     1 & 1 & * \\
                     1 & 1 & * \\
                     * & * & 1 \\
                     \end{array}
                 \right)$, and
                 $\left(\begin{array}{ccc}
                     1 & 1 & * \\
                     1 & * & 1 \\
                     * & * & 1 \\
                     \end{array}
                 \right)$.
By considering the possible choices of $Q_1$ and $Q_2$, noting that the phases in $Q_1$ and $Q_2$ that may affect the elements of $A$ can be assumed to be third roots of unity, it can be found that it is impossible to make the block $B$ have five identical elements in the same five positions within the block. This excludes case (a.2).

Thus, the case (b) holds. Let $m\in\{1,2,3\}$. Suppose the first column of $S$ is mapped by $Q_2$ to the $p$-th column of $T$. Let $q=p-3$ if $p>3$, otherwise let $q=p+3$. Suppose the preimage of the $q$-th column of $T$ under the action of $Q_2$ is the $r$-th column of $s$. Then $2\le r\le 6$.
All columns of $S$ except the dephased column have three distinct elements $1,\o,\o^2$, each appearing twice. Thus the $p$-th and the $q$-th columns of $T$ cannot be proportional to each other. This fact combined with the requirement of case (b) imply that $A$ is proportional to $D$, and $B$ is proportional to $C$, while $A$ and $B$ are linearly independent. In the following we assume without loss of generality that $p\le 3$, so the first three elements of the $p$-th column of $T$ is in $A$. The first column and the $r$-th column of $S$ satisfy that there are phases $\alpha,\beta,\gamma$ such that when the last $3$ elements of the first column and the $r$-th column of $S$ are multiplied by them, the two column vectors satisfy the relations that the parts of the columns in $A$ and $D$ (after a permutation of the columns caused by $Q_2$) are proportional to each other, and the parts of the columns in $B$ and $C$ are proportional to each other. After inspecting the possible values of $\alpha,\beta,\gamma$, noting that $\alpha$ can always be assumed to be $1$ while $\beta$ and $\gamma$ can be assumed to be third roots of unity, we find that the condition in the previous sentence cannot be satisfied, no matter what $r$ is. This implies that $S$ cannot be of min-Schmidt rank two. Thus the min-Schmidt rank of $S$ is exactly three.

(ix) We first prove for the case of the Fourier family $F_6^{(2)}$. Assume $U$ is a CHM of the form in \eqref{eq:ufourier1}.
Let $V=Q U Q^\dag$, where
\bea\label{eq:ufourier2}
Q=\left(
                   \begin{array}{cccccc}
                     1 & 0 & 0 & 0 & 0 & 0 \\
                     0 & 0 & 1 & 0 & 0 & 0 \\
                     0 & 0 & 0 & 0 & 1 & 0 \\
                     0 & 1 & 0 & 0 & 0 & 0 \\
                     0 & 0 & 0 & 1 & 0 & 0 \\
                     0 & 0 & 0 & 0 & 0 & 1 \\
                     \end{array}
                 \right).
\eea
The matrix $V$ contains four $3\times 3$ blocks. The upper two $3\times 3$ blocks of $V$ are equal, and the lower two $3\times 3$ blocks only differ by a minus sign. Hence $V$ is of Schmidt rank at most two. Thus $U$ is of min-Schmidt rank at most two. Any CHM in $F_6^{(2)}$ is equivalent to $U$ under complex permutation matrices, thus any CHM in $F_6^{(2)}$ is of min-Schmidt rank at most two.

The case of the Fourier transposed family is similar.
This completes the proof. \epf

\section{The proof of Lemma~\ref{le:mub}}

\label{app:{le:mub}}

\bpf
The statement (i) is clear. The local equivalence in (i) is the same as that in \cite[p4]{mw12jpa135307}, which applies to only the product-vector MUBs in (iii).

We prove the first equivalence of statement (ii). Suppose that the CHM $A$ is a member of an MUB trio, and that $I,A,B,C$ are four MUB matrices, then $I,A^\dag,A^\dag B,A^\dag C$ are another four MUB matrices, so the latter three matrices form an MUB trio, thus $A^\dag$ is a member of an MUB trio. Next, the second equivalence of statement (ii) follows from the definition of MUB trios. Further the third equivalence of statement (ii) follows from the first and second equivalence. Finally, statement (ii.a) follows from the definition of MUB trios.

Next, statement (iii) is from \cite[Theorem 4]{mw12jpa135307}.
The first part of statement (iv) is implied by the second paragraph below Eq. (22) of \cite{mw12jpa102001} together with the argument in \cite{jmm09}. If some two of four MUBs in $\bbC^6$ contain more than ten product vectors, then (i) together with the case $k=5$ of Lemma~\ref{le:la}~(ix) imply that the two MUBs are all product vectors. It is a contradiction with the first part of (iv). On the other hand the implication from the second statement of (iv) to the first statement of (iv) is clear. So we have proved the equivalence between the two statements.

Next we prove the statement (v) with Eqs. \eqref{eq:3x3subunitary}-\eqref{eq:2x2subunitary+2x2subsingular}. For each equation we shall assume that the $Y_j$ is a member of some MUB trio. Using the unitary equivalence in Definition \ref{df:mub} (i) and (iii), the matrix
\bea
\label{pyjq}
P Y_j Q
\eea
is still a member of some MUB trio for any order-six complex permutation matrices $P$ and $Q$. By regarding $PY_jQ$ as the new $Y_j$, we can place any submatrix of $Y_j$ in the upper-left corner of $Y_j$.

For (v) with \eqref{eq:3x3subunitary}, we assume that the order-three submatrix $V_0$ of the partitioned order-six CHM $Y_1=\left(
                   \begin{array}{cccccc}
                     V_0 & V_1 \\
                     V_2 & V_3 \\
                     \end{array}
                 \right)$ is a subunitary matrix.
By computation one can show that $\sqrt2 V_0$ and $\sqrt2 V_2$ are both order-three unitary matrices.
It follows from Definition  \ref{df:mub} (i) that there are four MUBs which contain two members
\bea
\label{eq:sqrt2v0}
\left(
                   \begin{array}{cccccc}
                     \sqrt2 V_0^\dg & 0 \\
                     0 & \sqrt2 V_2^\dg  \\
                     \end{array}
                 \right),
~~~~
\left(
                   \begin{array}{cccccc}
                     I_3/\sqrt2 & \sqrt2 V_0^\dg V_1 \\
                     I_3/\sqrt2 & \sqrt2 V_2^\dg V_3 \\
                     \end{array}
                 \right).
\eea
Since the second matrix is unitary, we obtain that $V_0^\dg V_1=-V_2^\dg V_3$.
So both matrices in \eqref{eq:sqrt2v0} represent product-vector bases. It is a contradiction with assertion (iv). So $Y_1$ is not a member of any MUB trio.

For (v) with \eqref{eq:3x2}, we assume that the upper left $3\times 2$ submatrix of $Y_2=[a_{ij}]$ has rank one and all elements $1/\sqrt6$, and all elements of the first column vector of $Y_2$ are also $1/\sqrt6$. Since $Y_2$ is an order-six CHM, we obtain that $a_{i2}=-1/\sqrt6$ for $i=4,5,6$.
By the same reason we obtain that the vector $(a_{1j},a_{2j},a_{3j})$ for $j=3,4,5,6$ is orthogonal to the vector $(1,1,1)$. So $(a_{1j},a_{2j},a_{3j})$ is proportional to $(1,\o,\o^2)$ or $(1,\o^2,\o)$ by Lemma \ref{le:la} (i). It follows from Lemma \ref{le:chm} (iii) that exactly two of
$(a_{1j},a_{2j},a_{3j})$ for $j=3,4,5,6$ are proportional to $(1,\o,\o^2)$ and exactly two of them are proportional to $(1,\o^2,\o)$. So the three upper rows of $Y_2$ contains an order-three subunitary matrix. The CHM $Y_2$ is not a member of any MUB trio by (v) with \eqref{eq:3x3subunitary}.

For (v) with \eqref{eq:3x31ortho2}, we assume that the partitioned order-six CHM $Y_3=\left(
                   \begin{array}{cccccc}
                     V_0 & V_1 \\
                     V_2 & V_3 \\
                     \end{array}
                 \right)$ has the order-three submatrix $V_0$, and
the first column and row vectors of $Y_3$ have all elements $1/\sqrt6$. Further the first column vector of $V_0$ is orthogonal to another two column vectors of $V_0$. They are ${1\over\sqrt6}(1,\o,\o^2)^T$ or ${1\over\sqrt6}(1,\o^2,\o)^T$ by Lemma \ref{le:la} (i). These two column vectors must be the same, from (v) with \eqref{eq:3x3subunitary}. The CHM $Y_3$ is not a member of any MUB trio by (v) with \eqref{eq:3x2}.

For (v) with \eqref{eq:3prod}, we assume that the leftmost three column vectors of $Y_4$ are all product vectors $\ket{a_j,b_j}$ for $j=1,2,3$, and the modulus of the elements of $\ket{a_j}$ and $\ket{b_j}$ are respectively $1/\sqrt2$ and $1/\sqrt3$.
If the $\ket{a_j}$ are pairwise non-orthogonal, then the $\ket{b_j}$ are pairwise orthogonal. So $Y_4$ contains an order-three subunitary matrix. This is a contradiction with (v) with \eqref{eq:3x3subunitary} because $Y_4$ is a member of some MUB trio. So two of $\ket{a_j},j=1,2,3$ are orthogonal. If the remaining vector is proportional to neither of them, then $Y_4$ contains an order-three submatrix whose one column vector is orthogonal to the other two column vectors of the submatrix. This is a contradiction with (v) with \eqref{eq:3x31ortho2}. So two of $\ket{a_j},j=1,2,3$ form an orthonormal basis in $\bbC^2$ and the third vector is proportional to one of them. Using suitable $P$ and $Q$ in \eqref{pyjq} we may assume that $\ket{a_1}=\ket{a_2}=(1,1)^T/\sqrt2$, $\ket{a_3}=(1,-1)^T/\sqrt2$, $\ket{b_1}=(1,\o,\o^2)^T/\sqrt3$, $\ket{b_2}=(1,\o^2,\o)^T/\sqrt3$, and the fourth column vector of $Y_4$ is denoted as $(1,a_{24},a_{34},a_{44},a_{54},a_{64})^T/\sqrt6$ with $\abs{a_{j4}}=1$. Since the first, second and fourth column vectors of $Y_4$ are pairwise orthogonal, we obtain that $(1+a_{44},a_{24}+a_{54},a_{34}+a_{64})$ is orthogonal to $\ket{b_1}$ and $\ket{b_2}$. Hence $(1+a_{44},a_{24}+a_{54},a_{34}+a_{64})\propto (1,1,1)$. We have
\bea
\label{eq:1+a44}
1+a_{44}=a_{24}+a_{54}=a_{34}+a_{64}.
\eea
If they are zero then the fourth column vector of $Y_4$ is a product vector $\ket{a_4,b_4}=(1,-1)^T/\sqrt2\ox(1,a_{24},a_{34})^T/\sqrt3$. Since $\ket{a_3,b_3}\perp\ket{a_4,b_4}$ and $\ket{a_3}=\ket{a_4}$, we have $\ket{b_3}\perp\ket{b_4}$. Let $\ket{g}\perp\ket{b_3},\ket{b_4}$. Since $Y_4$ is an order-six CHM, the rightmost two column vectors of $Y_4$ are in the span of $(1,1)^T\ox(1,1,1)^T$ and $(1,-1)^T\ox\ket{g}$, and proportional to neither of them by (iv) and Lemma \ref{le:la} (iii). This condition and the fact that $Y_4$ has elements of modulus $1/\sqrt6$ imply that the elements of $\ket{g}$ are $\pm1/\sqrt6$.
So the rightmost two column vectors of $Y_4$ form a submatrix of $Y_4$ containing a submatrix of size $3\times2$ and rank one. This is a contradiction with (v) with \eqref{eq:3x2}.

It remains to investigate the case that \eqref{eq:1+a44} is nonzero. Since $\abs{a_{j4}}=1$, Lemma \ref{le:la} (iv) implies that the two 2-tuples $(a_{24},a_{54})$ and $(a_{34},a_{64})$ are equal to $(1,a_{44})$ or $(a_{44},1)$. So the fourth column vector of $Y_4$ is one of the following four vectors
\bea
\label{eq:1oversqrt6}
{1\over\sqrt6}
\left(
                   \begin{array}{cccccc}
                     1 & 1 & 1 & 1\\
                     1 & 1 & a_{44}' & a_{44}''\\
                     1 & a_{44} & 1 & a_{44}''\\
                     a_{44}''' & a_{44} & a_{44}' & a_{44}''\\
                     a_{44}''' & a_{44} & 1 & 1\\
                     a_{44}''' & 1 & a_{44}' & 1\\
                     \end{array}
                 \right),
\eea
where we have used the modulus-one complex numbers $a_{44},a'_{44},a''_{44},a'''_{44}$ to distinguish the vectors. Since $\ket{b_1}=(1,\o,\o^2)^T/\sqrt3$, $\ket{b_2}=(1,\o^2,\o)^T/\sqrt3$, the first vector in \eqref{eq:1oversqrt6} is excluded by (v) with \eqref{eq:3x3subunitary}.
The same reason implies that $a_{44},a_{44}',a_{44}''\ne1$.
We apply the above argument to obtain that the fifth and sixth column vectors of $Y_4$ are also equal to one of the rightmost three column vectors of \eqref{eq:1oversqrt6}. (v) with \eqref{eq:3x2} implies that the rightmost three column vectors of $Y_4$ are respectively equal to the rightmost three column vectors of \eqref{eq:1oversqrt6} up to a permutation of the columns. They are orthogonal to $\ket{a_3,b_3}$ because $Y_4$ is unitary. Since $\ket{a_3}=(1,1)^T/\sqrt2$, $\ket{b_3}$ is orthogonal to $(1,1,-1),(1,-1,1)$ and $(-1,1,1)$. This is a contradiction with the fact that $\ket{b_3}$ is nonzero. So $Y_4$ is not a member of any MUB trio.

For (v) with \eqref{eq:3x3singular}, first of all we investigate the case when $Y_5$ contains a singular matrix of order three. We assume that the partitioned order-six CHM $Y_5=\left(
                   \begin{array}{cccccc}
                     V_0 & V_1 \\
                     V_2 & V_3 \\
                     \end{array}
                 \right)$ has the singular order-three submatrix $V_0$,  and
the first column and row vectors of $Y_5$ have all elements $1/\sqrt6$. If the second column vector of $V_0$ has three distinct entries then the third column vector of $V_0$ is proportional to one of the first two column vectors of $V_0$ by Lemma \ref{le:van}. It is a contradiction with (v) with \eqref{eq:3x2}. So the second column vector of $V_0$ has exactly two distinct entries. Since $V_0$ is singular, the third column vector of $V_0$ also has exactly two distinct entries, and the same entries in the second and third column vectors of $V_0$ are in the same row of $V_0$. So $V_0$ contains a $2\times3$ submatrix of rank one. The CHM $Y_5$ is not a member of any MUB trio by (v) with \eqref{eq:3x2} and (ii).

For (v) with \eqref{eq:3x2real},
we assume that the first column vector of $Y_6$ has all entries $1/\sqrt6$, the upper left $3\times 2$ submatrix of $Y_6$ is real, all but the last entry $x$ of the submatrix are equal to $1/\sqrt6$, and $x=1/\sqrt6$ or $-1/\sqrt6$. The former case is excluded by  (v) with \eqref{eq:3x2}.
If $x=-1/\sqrt6$, then Lemma \ref{le:la} (iv) shows that the bottom three elements of the second column vector of $Y_6$ are $-1/\sqrt6$, $s/\sqrt6$ and $-s/\sqrt6$ with some complex number $s$ of modulus one. Up to equivalence the first two column vectors are $(1,1)^T/\sqrt2\ox(1,1,1)^T/\sqrt3$ and $(1,-1)^T/\sqrt2\ox(1,1,s)^T/\sqrt3$. They are orthogonal to the third column vector of $Y_6$. By computation we can show that the third and the sixth element of the vector are zero. It is a contradiction with the fact that $Y_6$ is a CHM.

For (v) with \eqref{eq:2prod=ab,ac}, we assume that the leftmost two column vectors of $Y_7$ are product vectors $\ket{a,b}$ and $\ket{a,c}$.
Using suitable $P$ and $Q$ in \eqref{pyjq} we may assume that $\ket{a}=(1,1)^T/\sqrt2$, $\ket{b}=(1,\o,\o^2)^T/\sqrt3$, $\ket{c}=(1,\o^2,\o)^T/\sqrt3$, and the third column vector of $Y_7$ is denoted as $(1,a_{23},a_{33},a_{43},a_{53},a_{63})^T/\sqrt6$ with $\abs{a_{j3}}=1$. Since the column vectors of $Y_7$ are pairwise orthogonal, we obtain that $(1+a_{43},a_{23}+a_{53},a_{33}+a_{63})$ is orthogonal to $\ket{b}$ and $\ket{c}$. Hence $(1+a_{43},a_{23}+a_{53},a_{33}+a_{63})\propto (1,1,1)$. We have
\bea
\label{eq:1+a43}
1+a_{43}=a_{23}+a_{53}=a_{33}+a_{63}.
\eea
If they are zero then $Y_7$ contains a real submatrix of size $2\times3$. The assertion follows from (v) with \eqref{eq:3x2real}. If they are nonzero, then
 $\abs{a_{j3}}=1$ and Lemma \ref{le:la} (iv) imply that the two 2-tuples $(a_{23},a_{53})$ and $(a_{33},a_{63})$ are equal to $(1,a_{43})$ or $(a_{43},1)$. So the third column vector of $Y_7$ is one of the following four vectors
\bea
\label{eq:1oversqrt6a43}
{1\over\sqrt6}
\left(
                   \begin{array}{cccccc}
                     1 & 1 & 1 & 1\\
                     1 & 1 & a_{43}' & a_{43}''\\
                     1 & a_{43} & 1 & a_{43}''\\
                     a_{43}''' & a_{43} & a_{43}' & a_{43}''\\
                     a_{43}''' & a_{43} & 1 & 1\\
                     a_{43}''' & 1 & a_{43}' & 1\\
                     \end{array}
                 \right),
\eea
where we have used the modulus-one complex numbers $a_{44},a'_{44},a''_{44},a'''_{44}$ to distinguish the vectors. Since $\ket{b}=(1,\o,\o^2)^T/\sqrt3$, $\ket{c}=(1,\o^2,\o)^T/\sqrt3$, the first vector in \eqref{eq:1oversqrt6a43} is excluded by (v) with \eqref{eq:3x3subunitary}.
We apply the above argument to obtain that the rightmost four column vectors of $Y_7$ are equal to one of the rightmost three column vectors of \eqref{eq:1oversqrt6a43}. Hence two of the rightmost four column vectors of $Y_7$ are the same, and it is a contradiction with the fact that $Y_7$ is a CHM.
So $Y_7$ is not a member of any MUB trio.

For (v) with \eqref{eq:adjoint}, up to equivalence we may assume that $Y_8$ is in the dephased form and \eqref{eq:y8} is the upper left corner of $Y_8$. It follows from \cite[Lemma 2.7.]{nicoara08} (a) that two of $x,y,z$ are equal. We prove the assertion for $x=y$, and one can similarly prove the assertion for $x=z$ or $y=z$. Denote the last two elements on the second and third rows of $Y_8$ respectively as $u,v$ and $s,t$. The orthogonality of the first three rows of $Y_8$ implies that
\bea
2+x+x^*=-(u+v),
\\
2+x+z^*=-(s^*+t^*),
\\
1+2x^*+xz=-(u^*s+v^*t).
\eea
Then Lemma \ref{le:la} (vi) implies that
\bea
\label{eq:2+x}
(2+x+x^*)(2+x+z^*)(1+2x^*+x z)\in \bbR.
\eea
If $x=-1$ then $Y_8$ contains a real matrix of size $3\times2$. The assertion follows from (v) with \eqref{eq:3x2real}. If $x\ne-1$ then $2+x+x^*>0$. \eqref{eq:2+x} implies that $(2+x+z^*)(1+2x^*+xz)\in \bbR$. We have
$
(2+x+z^*)(1+2x^*+xz)
=
2+4x^*+2xz+x+2+x^2z+z^*+2x^*z^*+x\in \bbR.
$
Hence
$
x(-2+xz+x^*z^*)\in\bbR.
$
We have $x\in\bbR$ or $xz=1$. If $x\in\bbR$ then
the assertion follows from (v) with \eqref{eq:3x2real}. On the other hand if $xz=1$ then the orthogonality of the first two column vectors of $Y_8$ implies that
the last two elements of the second column vector of $Y_8$ are complex conjugates of each other. Let them be $a$ and $a^*$. The first four column vectors of $Y_8$ can be written as
\bea
\label{eq:y8a}
{1\over\sqrt6}
\left(
                   \begin{array}{cccccc}
                     1 &  1 & 1 & 1  \\
                     1 &  1 & x & x^* \\
                     1 &  x^* & 1 & x^* \\
                     1 &  x & x & 1 \\
                     1 &  a & b & d \\
                     1 &  a^* & c & e \\
                     \end{array}
                 \right),
\eea
where $b,c,d,e$ are complex numbers of modulus one. The orthogonality of column vector 1, 3 and 4 of \eqref{eq:y8a} implies that
$
2+2x^*=-b^*-c^*=-d-e.
$
The orthogonality of column 3 and 4 of \eqref{eq:y8a} implies that
$
(1+x^*)^2+b^*d+c^*e=0.
$ The two equations imply that
\bea
\label{eq:d+e}
(d+e)^2+4b^*d+4c^*e=0.
\eea
Applying Lemma \ref{le:la} (iv) to $-b^*-c^*=-d-e$, we obtain three cases: (i) $b=-c$, (ii) $b^*=e,c^*=d$, and (iii) $b^*=d,c^*=e$. In case (i), the orthogonality of column vector 1 and 3 of \eqref{eq:y8a} implies that $x=-1$. So $Y_8$ contains a real matrix of size $3\times2$. The assertion follows from (v) with \eqref{eq:3x2real}. In case (ii), \eqref{eq:d+e} implies that $\abs{d}\ne\abs{e}$. It is a contradiction with the fact that $\abs{d}=\abs{e}=1$. So case (ii) is excluded. In case (iii), the orthogonality of column vector 2, 3 and 4 of \eqref{eq:y8a} implies that $a^*d^*+ae^*=ad^*+a^*e^*$. Hence $a=a^*$ or $d=e$. If $a=a^*$ then the assertion follows from (v) with \eqref{eq:3x2real}. If $d=e$ then $b=c$. So $Y_8$ contains a $2\times3$ real matrix up to equivalence. So $Y_8$ is not a member of any MUB trio by assertion (ii) and (v) with \eqref{eq:3x2real}.

For (v) with \eqref{eq:2x2subunitary+2x2subsingular}, suppose $Y_9$ is a member of some MUB trio. It follows from Lemma \ref{le:chm} (vi) that $Y_9$ is equivalent to the CHM
$
{1\over\sqrt6}\left(
                   \begin{array}{cccccc}
                     A_{11} & A_{12} & A_{13}  \\
                     A_{21} & A_{22} & A_{23}  \\
                     A_{31} & A_{32} & A_{33}  \\
                     \end{array}
                 \right),
$
where any $A_{ij}$ is an order-two subunitary matrix and
\bea
\label{eq:y9-2}
\left(
                   \begin{array}{cccccc}
                     A_{11} \\
                     A_{21} \\
                     A_{31} \\
                     \end{array}
                 \right)
=\left(
                   \begin{array}{cccccc}
                     1 & 1 \\
                     1 & -1 \\
                     1 & a \\
                     1 & -a \\
                     1 & b \\
                     1 & -b \\
                     \end{array}
                 \right).
\eea
If $Y_9$ has exactly two columns containing an order-two subunitary matrix $V$ and an order-two singular matrix $W$, then we may assume that the two columns are \eqref{eq:y9-2}. Then we have $a,b\in\{1,-1\}$ or $a\in\{b,-b\}$. Either case is excluded by (v) with \eqref{eq:3x2real}.

For (v) with \eqref{eq:4x3}, up to equivalence we may assume that the upper left submatrix $V$ of size $4\times3$ of $Y_{10}$ satisfies the hypothesis, i.e., the first column vector of $V$ is orthogonal to the remaining two column vectors of $V$. Let $W$ be the lower left submatrix in $Y_{10}$ of size $2\times3$ that is below the submatrix $V$. Since $Y_{10}$ is a CHM, the above facts imply that the first column vector of $W$ is orthogonal to the  remaining two column vectors of $W$. Up to equivalence We have $W=\left(
                   \begin{array}{cccccc}
                     1 & 1 & 1 \\
                     1 & -1 & -1 \\
                     \end{array}
                 \right)$. It follows from (ii) and (v) with \eqref{eq:3x2real} that $Y_{10}$ is not a member of any MUB trio.

For (v) with \eqref{eq:3x3v}, since $Y_{11}$ is an order-six CHM, the submatrices $V$ and $D_1VD_2$ both consist of elements of modulus $1/\sqrt6$. Since $D_1$ and $D_2$ are GPMs, they are both complex permutation matrices up to global factors. If $Y_{11}$ is a member of some MUB trio, then so is $(D_1^\dg \op I_3) Y_{11}$. This is a matrix of expression
\bea
\label{eq:v2y}
\left(
                   \begin{array}{cccccc}
                     v_2 & y v_3 & z v_1 \\
                     v_1 & v_2 & v_3 \\
                     \end{array}
                 \right),
 \eea
where $v_1,v_2$ and $v_3$ are all 3-dimensional vectors of modulus $1/\sqrt6$, and $y$ and $z$ complex numbers. The orthogonality between the column vectors implies that $v_1,v_2$ and $v_3$ are pairwise orthogonal. It is a contradiction with (v) with \eqref{eq:3x3subunitary}. So $Y_{11}$ is not a member of any MUB trio.

Next we prove (v) with \eqref{eq:d1vd2}. If an MUB trio contains $\bigg[
U\ox I_3\bigg]
\cdot
\left(
                   \begin{array}{cccccc}
                     V_1 &  0 \\
                     0 &  D_1V_1D_2 \\
                     \end{array}
                 \right)
\cdot
\bigg[
X \ox I_3\bigg]$ then Definition \ref{df:mub} (i) implies the existence of four MUBs containing $(I_3\op D_1^\dg)\cdot\bigg[
U^\dg\ox I_3\bigg]$ and $\left(
                   \begin{array}{cccccc}
                     V_1 &  0 \\
                     0 &  V_1D_2 \\
                     \end{array}
                 \right)
\cdot \bigg[ X\ox I_3 \bigg]$. They are both product-vector bases. It is a contradiction with assertion (iv). So we have proved (v) with \eqref{eq:d1vd2}.

Finally we prove (v) with \eqref{eq:xi3}.
In case (1), denote the $3\times 6$ matrix formed by the top three rows of the matrix in \eqref{eq:xi3} as $T$. It can be shown that $T$ contains a singular order-three submatrix. To see this, note that there may be two cases depending on the location of zero elements in $V_3$ and $W_3$ in \eqref{eq:xi3}: one is that the same rows in $V_3$ and $W_3$ contains two zero elements, in which case there are two columns of $T$ that are proportional to each other; and in the remaining case, three columns of $T$ form a linearly dependent set, where one column is from the first three or the last three columns of $T$, and the remaining two columns are from the other three columns of $T$. This is a contradiction with the claim about the matrix $Y_5$ in assertion (v). We have proved (v) with \eqref{eq:xi3} for case (1). In case (2), the CHM contains a singular order-three submatrix. It is excluded by (v) with \eqref{eq:3x3singular}. In case (3), if $D_3\propto D_4$ then both of them are proportional to the identity matrix. So the CHM contains an order-three subunitary submatrix.  It is excluded by (v) with \eqref{eq:3x3subunitary}. In case (4) similarly to case (1), one can show that the top three rows contain three column vectors that are linearly dependent. We have proved (v) with \eqref{eq:xi3} for case (4).
This completes the proof.
\epf

\section{The proof of Lemma~\ref{le:sr2}}

\label{app:{le:sr2}}

\bpf
(i) Any Schmidt-rank-one order-six CHM can be written as $U\ox V$ where $U$ and $V$ are respectively an order-two and order-three CHM.  It is known that $U$ is equivalent to
$
{1\over\sqrt2}
\left(
                   \begin{array}{cccccc}
                     1 &  1 \\
                     1 &  -1 \\
                     \end{array}
                 \right),
$
and $V$ is equivalent to
\bea\label{eq:schm3}
{1\over\sqrt3}
\left(
                   \begin{array}{cccccc}
                     1 &  1 & 1 \\
                     1 &  \o & \o^2 \\
                     1 &  \o^2 & \o \\
                     \end{array}
                 \right),
\eea
see \cite{mub09}. So the assertion holds. Note that The CHM $\bbH_1$ stands for a product-vector basis in $\bbC^6$.

(ii) Any Schmidt-rank-two bipartite unitary is a controlled unitary \cite{cy13}. So any Schmidt-rank-two order-six CHM $Y$ can be written as $(S\ox I_3)(T\op U) (X\ox I_3)$ where $S$ and $X$ are order-two unitary matrices, and $T$ and $U$ are order-three unitary matrices. We can find order-two complex permutation matrices $D_1,D_2$ and $D_3$ such that $SD_1=\left(
                   \begin{array}{cccccc}
                     \cos\a &  \sin\a \\
                     e^{i\g}\sin\a &  -e^{i\g}\cos\a \\
                     \end{array}
                 \right)$, $D_2XD_3=\left(
                   \begin{array}{cccccc}
                     \cos\b &  \sin\b \\
                     \sin\b &  -\cos\b \\
                     \end{array}
                 \right)$, $\a,\b\in[0,\p/4]$, $\g\in[0,2\p]$ and $(D_1^\dg\ox I_3)(T \op U)(D_2^\dg\ox I_3)=V\op W$. Since $Y$ has Schmidt rank two and local unitaries does not change the Schmidt rank of $Y$, we obtain that $V$ and $W$ are linearly independent. So we have obtained \eqref{eq:v=w} and
\bea
\label{eq:y}
Y=
\left(
                   \begin{array}{cccccc}
                     (\cos\a\cos\b) V + (\sin\a\sin\b) W & (\cos\a\sin\b) V - (\sin\a\cos\b) W \\
                    e^{i\g} (\sin\a\cos\b) V -e^{i\g} (\cos\a\sin\b) W &e^{i\g} (\sin\a\sin\b) V + e^{i\g}(\cos\a\cos\b) W \\
                     \end{array}
                 \right).
\eea
Let $v$ and $w$ be any two entries of the same position in $V$ and $W$, respectively. Since $Y$ is an order-six CHM we have
\bea
&&
\label{eq:cc+ss}
\abs{(\cos\a\cos\b) v + (\sin\a\sin\b) w}
\\
&=&
\label{eq:cs-sc}
\abs{(\cos\a\sin\b) v - (\sin\a\cos\b) w}
\\
&=&
\label{eq:sc-cs}
\abs{(\sin\a\cos\b) v - (\cos\a\sin\b) w}
\\
&=&
\label{eq:ss+cc}
\abs{(\sin\a\sin\b) v + (\cos\a\cos\b) w}
\\
&=&
\label{eq:1sqrt6}
{1\over\sqrt6}.
\eea
The equality between \eqref{eq:cc+ss}, \eqref{eq:cs-sc} and \eqref{eq:1sqrt6} is equivalent to
\bea
\label{eq:cv+sw}
&&
\abs{(\cos\a\cos\b) v + (\sin\a\sin\b) w}^2
+
\abs{(\cos\a\sin\b) v - (\sin\a\cos\b) w}^2
\notag\\
&=&
(\cos\a)^2 \abs{v}^2+(\sin\a)^2 \abs{w}^2
\notag\\
&=&1/3,
\eea
and
\bea
\label{eq:cv-sw}
&&
\abs{(\cos\a\cos\b) v + (\sin\a\sin\b) w}^2
-
\abs{(\cos\a\sin\b) v - (\sin\a\cos\b) w}^2
\notag\\
\label{eq:cv-sw2}
&=&
((\cos\a)^2 \abs{v}^2-(\sin\a)^2 \abs{w}^2)\cos2\b
+({vw^*+v^*w\over2})\sin2\a\sin2\b
\notag\\
&=&0.
\eea
The equality between \eqref{eq:cc+ss} and \eqref{eq:ss+cc}, and the equality between \eqref{eq:cs-sc} and \eqref{eq:sc-cs} is equivalent to
\bea
\label{eq:cc2-ss2}
(\abs{v}^2-\abs{w}^2)\cos2\a=
(\abs{v}^2-\abs{w}^2)\cos2\b=0.
\eea
Then \eqref{eq:cv+sw} and \eqref{eq:cc2-ss2} imply \eqref{eq:sr2-2}. Next, \eqref{eq:sr2-2} and \eqref{eq:cc2-ss2} imply \eqref{eq:sr2-3} and \eqref{eq:sr2-4}. Third, \eqref{eq:sr2-2}, \eqref{eq:cv-sw2} and \eqref{eq:cc2-ss2} imply \eqref{eq:sr2-1}. By applying $vw^*+v^*w\ge-\abs{v}^2-\abs{w}^2$ to \eqref{eq:sr2-1} and \eqref{eq:sr2-2}, we obtain $\a+\b\ge\p/4$.
Thus \eqref{eq:sr2-0} is proved.
Next, \eqref{eq:sr2-0}-\eqref{eq:sr2-4} imply (ii.a) and (ii.b).

(ii.c) Since $\a+\b\ge\p/4$ in \eqref{eq:sr2-0}, it suffices to exclude the equation $\a+\b=\p/4$. Suppose it holds. The case $\a\b=0$ has been excluded by Lemma \ref{le:mub} (iv). So we may assume that $\a\b\ne0$. It follows from \eqref{eq:sr2-2}-\eqref{eq:sr2-4} that $\abs{v_{jk}}=\abs{w_{jk}}=1/\sqrt3$. It follows from \eqref{eq:sr2-1} that $v_{jk}=-w_{jk}$, and thus $X$ has Schmidt rank one. It is a contradiction with the fact that $X$ has Schmidt rank two. So $\a+\b>\p/4$.

(ii.d) Suppose $(\a,\b)=(\p/4,\p/4)$. Assertion (ii.b) implies that $v_{jk}^*w_{jk}+v_{jk}w_{jk}^*=0$. Then $v_{jk}= i p_{jk} w_{jk}$ where $p_{jk}=1$ or $-1$ for any $j,k$. Assertion (ii.b) also implies that $\abs{v_{jk}}^2+\abs{w_{jk}}^2=2/3$. Since $V$ and $W$ are both unitary, each of them contains at most one zero entry in the same position of them. We have
\bea
V=D_1\cdot
\bigg[
i\left(
                   \begin{array}{cccccc}
                     w_{11} &  w_{12} & w_{13} \\
                     w_{21} &  q_{22} w_{22} & q_{23} w_{23} \\
                     w_{31} &  q_{32} w_{32} & q_{33} w_{33} \\
                     \end{array}
                 \right)
\bigg]
\cdot D_2,
\eea
where $D_1$ and $D_2$ are diagonal real unitary matrices, and $q_{jk}=\pm1$. It follows from Lemma \ref{le:la} (ii) that $V=i D_1' W D_2'$ where $D_1'$ and $D_2'$ are diagonal real unitary matrices. It follows from \eqref{eq:d1vd2} in Lemma \ref{le:mub} (v) that $\bbH_2(\a,\b,\g,V,W)$ is not a member of MUB trio.

It remains to prove the assertion when $(\a,\b)\ne(\p/4,\p/4)$. Suppose $\bbH_2(\a,\b,\g,V,W)$ is a member of MUB trio. Since $(\a,\b)\ne(\p/4,\p/4)$, assertion (ii.a) implies that $V$ and $W$ are both CHMs.
Lemma \ref{le:chm} (i) implies that $V=D_1XD_2$ and $W=D_3YD_4$, where $D_1,\cdots,D_4$ are all diagonal unitaries, $X$ and $Y$ are ${1\over\sqrt3}\left(
                   \begin{array}{cccccc}
                     1 &  1 & 1 \\
                     1 &  \o & \o^2 \\
                     1 &  \o^2 & \o \\
                     \end{array}
                 \right)$ and ${1\over\sqrt3}\left(
                   \begin{array}{cccccc}
                     1 &  1 & 1 \\
                     1 &  \o^2 & \o \\
                     1 &  \o & \o^2 \\
                     \end{array}
                 \right)$. Further $X\ne Y$ follows from \eqref{eq:d1vd2} in Lemma \ref{le:mub} (v).
Hence $\bbH_2(\a,\b,\g,V,W)$ is locally equivalent to the CHM $\bbH_2(\a,\b,\g,V',W')$ where $V'={1\over\sqrt3}\left(
                   \begin{array}{cccccc}
                     1 &  1 & 1 \\
                     1 &  \o & \o^2 \\
                     1 &  \o^2 & \o \\
                     \end{array}
                 \right)$, $W'=\diag(e^{i\g_1},e^{i\g_2},e^{i\g_3}) \cdot \bigg[{1\over\sqrt3}\left(
                   \begin{array}{cccccc}
                     1 &  1 & 1 \\
                     1 &  \o^2 & \o \\
                     1 &  \o & \o^2 \\
                     \end{array}
                 \right)\bigg] \cdot \diag(1,e^{i\d_2},e^{i\d_3})$ and $\g_1,\g_2,\g_3,\d_2,\d_3\in[0,2\p)$. Applying \eqref{eq:sr2-1} to $\bbH_2(\a,\b,\g,V',W')$, we obtain that the nine real numbers
\bea
\label{eq:g123}
\g_1,\g_2,\g_3,\g_1+\d_2,\g_2+\d_2+2\p/3,\g_3+\d_2+4\p/3,\g_1+\d_3,\g_2+\d_3+4\p/3,\g_3+\d_3+2\p/3
\eea
are equal to $2m\p\pm\arccos(-{1\over3}\cot2\a\cot2\b)$ with integers $m$.
So two of $\g_1,\g_2,\g_3$ are the same.

It follows from \eqref{eq:g123} that two of $\g_1+\d_2,\g_2+\d_2+2\p/3,\g_3+\d_2+4\p/3$ has difference $2\p/3$ or $4\p/3$. Hence
\bea
\label{eq:2p/3}
2\p/3=2n\p\pm2\arccos(-{1\over3}\cot2\a\cot2\b)
\eea
for some integer $n$ and one of the two signs $\pm$ holds.
On the other hand Eq. \eqref{eq:sr2-0} and assertion (ii.c) imply that $\cot2\a\cot2\b\in[0,1]$. Hence $\arccos(-{1\over3}\cot2\a\cot2\b)\in[\p/2,\arccos(-{1\over3})]$ where $\arccos(-{1\over3})\approx1.91$. So \eqref{eq:2p/3} does not hold. We have proved  that $\bbH_2(\a,\b,\g,V,W)$ is not a member of MUB trio.

(iii) Suppose $U$ is a Schmidt-rank-three order-six CHM. It is known that $U$ is a controlled unitary from the B side \cite{cy14}. We have $U=\sum^3_{j=1} (I_2\ox V) (U_j \ox \proj{j}) (I_2 \ox W)$ for some order-three unitary matrices $V$ and $W$ and some order-two unitary matrices $U_j$. By absorbing the phases of the first entries of the $U_j$'s into $V$, we may assume that these entries are $\cos\a_j$, $\a_j\in[0,\p/2]$ for $j=1,2,3$. The unitarity of $U_j$ then gives rise to the expression of $\bbH_3(\a_1,\b_1,\g_1,\a_2,\b_2,\g_2,\a_3,\b_3,\g_3,V,W)$ in the assertion.

It remains to prove the assertion that we can make the first column vector of $W$ consist of nonnegative and real elements. Let the vector be $(a e^{i b},c e^{i d}, f e^{i g})^T$ where $a,c,f\ge0$ and $b,d,g$ are real. Since the matrix $I_2\ox\diag(e^{i b}, e^{i d}, e^{i g})$ commutes with the middle matrix of \eqref{eq:sr3}, we can replace $V$ by $V\cdot\diag(e^{i b}, e^{i d}, e^{i g})$. So the assertion holds.

(iii.a) Suppose one of $\a_1,\a_2,\a_3$ is equal to $0$ or $\p/2$. Then one of the four order-three submatrices forming the matrix $\bbH_3:=\bbH_3(\a_1,\b_1,\g_1,\a_2,\b_2,\g_2,\a_3,\b_3,\g_3,V,W)$ in \eqref{eq:sr3} is singular. So the matrix cannot be a member of any MUB trio by the matrix $Y_5$ in Lemma \ref{le:mub} (v). It gives us a contradiction and thus none of $\a_1,\a_2,\a_3$ is not equal to $0$ and $\p/2$.

This completes the proof.
\epf

\end{document}